\documentclass[prb,aps,twocolumn,showpacs,floatfix,eqsecnum]{revtex4}
\usepackage{graphicx}

\newcommand{\RR} {{\bf R}}
\newcommand{\N} {{\cal N}}
\newcommand{\kk} {{\bf k}}
\newcommand{\rr} {{\bf r}}
\newcommand{\pp} {{\bf p}}
\newcommand{\E} {{\cal E}}
\newcommand{\sbf} {{s_{bf}}}
\newcommand{\qq} {{\bf q}}
\newcommand{\PP} {{\bf P}}
\newcommand{\gt} {{\tilde g}}
\newcommand{\GR} {{\cal G}}
\newcommand{\ph} {{\phi}}
\newcommand{\Tm} {{\cal T}}
\newcommand{\QA} {{Q_0}}
\newcommand{\QB} {{\bar Q}}
\newcommand{\NA} {{N_0}}
\newcommand{\NB} {{\bar N}}
\newcommand{\GB} {{\bar G}}
\newcommand{\GRB} {{\bar \GR}}
\newcommand{\phA} {{\ph_0}}
\newcommand{\gtA} {{\gt^0}}
\newcommand{\SSS} {{\cal S}}

\newcommand{\Tt} {{\tilde T}}
\newcommand{\ab} {{a^*}}

\newcommand{\T} {{\bf T}}

\begin{document}

\title{
Dilute limit of a strongly-interacting
model of spinless fermions and hardcore bosons on the square lattice
}

\author{ N. G.~Zhang$^*$ and C.~L.~Henley}
\affiliation{Department of Physics, 
Cornell University, Ithaca, New York 14853-2501}

\begin{abstract}
In our model, spinless fermions (or hardcore bosons) on a square lattice
hop to nearest neighbor sites, and also experience a hard-core repulsion 
at the nearest neighbor separation. 
This is the simplest model of correlated electrons and is
more tractable for exact diagonalization than the Hubbard model.
We study systematically the dilute limit of this model by a combination
of analytical and several numerical approaches:
the two-particle problem using lattice Green
functions and the t-matrix, the few-fermion problem using a modified
t-matrix (demonstrating that the interaction energy is
well captured by pairwise terms), 
and for bosons the fitting of the energy as a function
of density to Schick's analytical result for dilute hard disks. 
We present the first systematic study for a 
strongly-interacting lattice model of the t-matrix, 
which appears as the central object in older theories 
of the existence of a two-dimensional Fermi liquid
for dilute fermions with strong interactions. 
For our model, we can (Lanczos) diagonalize the $7\times 7$ 
system at all fillings and the $20\times 20$ system
with four particles, thus going far beyond
previous diagonalization works on the Hubbard model.

\end{abstract}

\maketitle

\section{introduction}

Since the discovery of the high-temperature superconductors
in 1986, there has been intense study of a number of 
two-dimensional models that are believed to model
the electronic properties of the CuO$_2$ plane
of the cuprate superconductors,
for example, the Hubbard model, the $t-J$ model,
and the Heisenberg model.\cite{Dagotto,Manou}
Two-dimensional quantum models with
short-range kinetic and interaction
terms are difficult to study. In one dimension,
there are exact solutions using the Bethe ansatz
and a host of related analytical techniques,\cite{LiebMattis} 
and there is a very 
accurate numerical method, the density-matrix
renormalization group (DMRG),\cite{dmrg} that
can be applied to large systems relatively
easily. In two dimensions, on the other hand,
there are few exact solutions (one famous nontrivial case is the
Hubbard model with one hole in a half-filled
background, the Nagaoka state\cite{Nagaoka}),
and current numerical methods are not
satisfactory (quantum Monte Carlo
is plagued by the negative sign problem\cite{Dagotto} at
low temperatures and at many fillings of interest 
and the DMRG in two dimensions\cite{2Ddmrg}
is still in early development stage).

The most reliable method for studying
complicated quantum systems is exact diagonalization,
which means enumerating all basis states and 
diagonalizing the resulting Hamiltonian matrix.
Of course, this method is computationally 
limited by the growth of the Hilbert space
which is in general exponential in the number
of particles and the lattice size.
The $4\times 4$ Hubbard model with 16 electrons,
8 spin-up and 8 spin-down, after 
reduction by particle conservation, translation,
and the symmetries of the square, has 
1,310,242 states in the largest matrix 
block,\cite{Lin} and can be diagonalized
using the well-known Lanczos method.
The Hubbard model 
has been diagonalized for the $4\times 4$ lattice
(see e.g., Ref.~\onlinecite{Fano}),
and at low filling (four electrons) for
$6\times 6$\cite{Bruus} with extensive employment
of symmetries.

\subsection{The spinless fermion model}

We have asked the question: {\it Is there
a model that contains the basic ingredient of short-range
hopping and interaction but is simpler, in the exact
diagonalization sense, than the Hubbard model?}
The answer is yes: we can neglect the spin.
We obtain the following Hamiltonian for {\it spinless}
fermions,
	\begin{equation}
	H=-t\sum_{\langle i,j \rangle}\left(
	c^\dagger_i c_j + c^\dagger_j c_i\right)
	+V\sum_{\langle i,j \rangle} \hat n_i \hat n_j,
	\label{eq-Ham}
	\end{equation}
where $c^\dagger_i$ and $c_i$ are spinless fermion creation 
and annihilation operators at site $i$, $\hat n_i=c^\dagger_i c_i$
the number operator, $t$ the nearest-neighbor hopping amplitude, and $V$
the nearest-neighbor interaction.
Note that with spinless fermions, there can be at the most
one particle per site; no on-site interaction 
(as that in the Hubbard model) is possible, and we have
included in our Hamiltonian nearest-neighbor repulsion.

The spinless fermion model, Eq.~(\ref{eq-Ham}), is a two-state
model, and the number of basis states for a $N$-site system
is $2^N$, which is a significant reduction from
the $4^N$ of the Hubbard model. We can further reduce
the number of basis states by taking the nearest-neighbor
interaction $V=+\infty$, i.e., infinite repulsion,
which excludes nearest neighbors, giving roughly
$2^{N/2}$ states.

The spinless fermion model with infinite repulsion 
Eq.~(\ref{eq-Ham}) contains a significant reduction of 
the Hilbert space. After using particle conservation and 
translation symmetry (but not point group symmetry), 
the largest matrix for the $7\times 7$ system 
has $1,906,532$ states (for 11 particles), and
we can therefore compute for 
all fillings the $7\times 7$ system whereas 
for the Hubbard model $4\times 4$ is basically the limit.
This of course means that for certain limits 
we can also go much further than the Hubbard
model, for example, we can handle four particles
on a $20\times 20$ lattice where
the number of basis states is $2,472,147$.
This extended capability with our model
has enabled us to obtain a number of results 
that are difficult to obtain with the Hubbard model.

An added feature of our model is that the basis
set for the spinless fermion problem is identical
to that for the hardcore boson problem, because with hardcore
repulsion, there can be at the most one boson at one site also.
Therefore, without computational difficulty, we 
can study numerically both the spinless fermion
and hardcore boson problem.

Spinless fermions can also be realized in experiments,
for example, the spin polarized $^3$He due to 
a strong magnetic field, or
ferro or ferri-magnetic electronic systems where
one spin-band is filled. The one-dimensional
spinless fermion model with
finite repulsion is solved exactly using Bethe ansatz.\cite{YangYang} 
The infinite-dimensional
problem is studied in Ref.~\onlinecite{Uhrig}.
A very different approach using the renormalization
group for fermions is done in Ref.~\onlinecite{Shankar}.
A Monte Carlo study of the two-dimensional model
at half-filling only and low temperatures
is in Ref.~\onlinecite{Guber}, which, dating back to 1985,
is one of the earliest quantum Monte Carlo simulations
for fermions. (It is no coincidence that they chose
the model with the smallest Hilbert space.)

Considering the tremendous effort that has been devoted to
the Hubbard model and the close resemblance of 
our model, Eq.~(\ref{eq-Ham}), to the Hubbard model,
it is surprising that works on this spinless model
have been rather sparse, though it has been 
commented that the spinless model offers considerable 
simplifications.\cite{VollFermi}

This paper is one of the two that we are publishing to study
systematically the two-dimensional spinless 
fermion and hardcore boson model with infinite nearest-neighbor
repulsion. The present paper focuses on the dilute limit, 
treating the problem of a few particles, 
and the other paper\cite{stripepaper} will focus on the dense limit, 
near half-filled,\cite{FN-halffilled} 
where stripes (that are holes lining up across the lattice)
are natural objects (see Ref.~\onlinecite{HenleyZhang}
for a condensed study of stripes in this model). 
We will use Lanczos exact diagonalization, 
exploiting the much-reduced Hilbert space of our model, 
and a number of analytical techniques, 
for example, in this paper, lattice Green functions 
and the t-matrix. One of the goals of these two papers 
is to advertise this model 
of spinless fermions to the strongly-correlated 
electron community, as we believe that it is
the simplest model of correlated fermions
and deserves more research effort and 
better understanding.

The prior work most comparable to ours may be the studies of four
spinless electrons in a $6\times 6$ lattice, with Coulomb repulsion, 
by Pichard {\it et al};~\cite{Pichard} their motivation was the Wigner crystal
melting and the competition of Coulomb interactions with 
Anderson localization when a disorder potential is turned on.

\subsection{The t-matrix}
\label{sec-introtmat}

At the dilute limit of our model,
the scattering t-matrix is of fundamental importance.
For two particles, we expect that, at least when the potential
$V$ is small, we can write
a perturbative equation for energy,
	\begin{equation}
	\label{eq-perturb}
	E=\E(\qq_1)+\E(\qq_2)+\Delta E(\qq_1,\qq_2),
	\end{equation}
which is to say that the exact interacting
energy of two particles is the noninteracting
energy $\E(\qq_1)+\E(\qq_2)$, for a pair
of momenta $\qq_1$ and $\qq_2$, plus
a correction term $\Delta E$ 
due to the interaction $V$. 
And with more than two particles, at least when the
particle density is low, we expect to have
	\begin{equation}
E=\sum_\qq \E(\qq)+\frac{1}{2}\sum_{\qq,\qq'} \Delta E(\qq,\qq').
	\label{eq-perturbmany}
	\end{equation}
Eq.~(\ref{eq-perturbmany}) is central in Fermi liquid theory, 
where it is justified by 
the so-called ``adiabatic continuation'' idea,
which says that interacting fermion states correspond
one-to-one to noninteracting ones as we slowly 
switch on a potential. 

In the boson case, because
many bosons can occupy one quantum mechanical state and
form a condensate, Eq.~(\ref{eq-perturbmany}) 
should be modified, but with only two bosons, 
we expect Eq.~(\ref{eq-perturb}) should be valid 
(in that the correction vanishes in the dilute limit). 
Eqs.~(\ref{eq-perturb}) and (\ref{eq-perturbmany})
are used when we look at a list of 
noninteracting energies and draw correspondences 
with the interacting energies, the energy shift being
packaged in the term $\Delta E$.

One possible objection to the above formulas (Eqs.~(\ref{eq-perturb})
and (\ref{eq-perturbmany})) is that they appear to be perturbative, 
yet the interaction potential in our problem
is infinitely strong, so the first-order (first Born approximation) scattering
amplitude, being proportional to the potential, is infinite too.
However, this singular potential scattering problem 
(e.g., hard-sphere interaction in 3D)
has been solved (see Ref.~\onlinecite{Landau}) by replacing
the potential with the so-called scattering length,
which is finite even when the potential is infinite.
As we review in Appendix~\ref{sec-physical}, 
a perturbation series 
(Born series) can be written down (that 
corresponds to a series of the so-called ladder diagrams)
and even though each term is proportional to
the potential, the sum of all terms (the t-matrix,
$\Delta E$ in Eqs.~(\ref{eq-perturb}) and (\ref{eq-perturbmany}))
is finite.

Because the t-matrix captures two-body interaction
effects, it is the centerpiece of dilute fermion
and boson calculations with strong interactions. 
Field-theoretical calculations in both three and two dimensions are
based on the ladder diagrams and the t-matrix.
See Fetter and Walecka\cite{Fetter}
for the 3D problem, Schick\cite{Schick} 
for the 2D boson problem and Bloom\cite{Bloom}
the 2D fermion problem. 
For lattice fermion problems, Kanamori\cite{Kanamori}
derived the t-matrix for a tight-binding model
that is essentially a Hubbard model
(this work is also described in Yosida\cite{Yosida}).
And in Ref.~\onlinecite{Mattis}, 
the t-matrix is worked out explicitly for the Hubbard model,
and Kanamori's result is obtained. 
Ref.~\onlinecite{Mattis} 
also evaluated the t-matrix for the dilute limit in 
three dimensions and obtained a functional dependence on 
particle density. 

Rudin and Mattis\cite{RudinMattis} 
used the t-matrix expression derived in 
Refs.~\onlinecite{Kanamori} and \onlinecite{Mattis}
and found upper and lower bounds of the fermion t-matrix
in two dimensions in terms of particle density.
Rudin and Mattis's result for the low-density limit
of the two-dimensional Hubbard model is
of the same functional form as Bloom's diagrammatical calculation
for the two-dimensional fermion hard disks.\cite{Bloom}
Since the discovery of high-temperature superconductors,
Bloom's calculation has received a lot of attention
because of its relevance to
the validity of the Fermi liquid description of
dilute fermions in two dimensions. There have been a number of 
works on the 2D dilute Fermi gas\cite{Engel1,Engel2,Engel3}
and on the dilute limit of 2D Hubbard model,\cite{FukuHubbard} 
all using the t-matrix, 
but these results have not been checked by 
numerical calculations.

In fact, we are not aware of a systematic study of 
the t-matrix for a lattice model. In this paper, we 
present the first such study for the two-particle problem
in Sec.~\ref{sec-tmat} (for bosons and fermions) 
and the few-fermion problem in Sec.~\ref{sec-tmat2}.
We check the t-matrix results with exact diagonalization
data and show that our t-matrix on a lattice
is the sum of the two-body scattering terms to
infinite order.

\subsection{Paper organization}

In this paper, we will study systematically the dilute
limit of our model Eq.~(\ref{eq-Ham}), focusing on the problem of a few 
particles. Our paper is divided into four parts.

In Sec.~\ref{sec-green}, the two-particle (boson and fermion) 
problem is studied. We formulated the two-particle 
Schrodinger equation using lattice Green functions,
employ some of its recursion relations to simplify
the problem, and obtain the two-boson ground state
energy in the large-lattice limit. 
Using the two-particle result, we then study the
problem of a few particles and obtain an expression
for ground state energy on a large lattice.

In Sec.~\ref{sec-tmat}, the two-particle problem
is then cast into a different form, emphasizing the scatterings
between the two particles. The result is the t-matrix, that is exact for the 
two-particle problem and contains all two-body scattering terms.
We will study the two-particle t-matrix in great detail, 
showing the differences between the boson 
and fermion cases, and demonstrating that the first t-matrix 
iteration is often a good approximation for fermion energy.
In Appendix~\ref{sec-physical}, we show explicitly
that the t-matrix we obtain is the sum total
of all two-body scattering terms.

The problem of a few fermions is taken up in Sec.~\ref{sec-tmat2}. 
First, the fermion shell effect is discussed 
and demonstrated from diagonalization, and
we show the difference for bosons and fermions.
We show the modifications
to the two-fermion t-matrix that enable us
to calculate energies for three, four, and five particles.
Using this t-matrix, we can compute the interaction
corrections to the noninteracting energy
and trace the change in the energy spectrum
from the nointeracting one to the interacting one.

Finally, in Sec.~\ref{sec-dilute},
we discuss the energy per particle curve for dilute bosons and fermions.
We have studied the two-dimensional results derived
by Schick\cite{Schick} for bosons and Bloom\cite{Bloom} for fermions
by fitting the data from diagonalization for a number of lattices.
Schick's result for dilute bosons is checked nicely,
and we explain that for spinless fermions in our model 
we will need the p-wave scattering term, which is not
included in Bloom's calculation.

In Appendix~\ref{sec-diag}, we discuss briefly our exact diagonalization 
computer program, which can handle arbitary periodic 
boundaries specified by two vectors on the 
square lattice and uses translation symmetry
to reduce the matrix size.

\section{The Two-Particle Problem}
\label{sec-green}

The two-particle problem has appeared in many different
contexts. The most familiar one is the hydrogen atom
problem in introductory quantum mechanics textbooks.
The two-magnon problem is closely related mathematically
to our two-particle problem, and it has been 
solved in arbitary dimensions for ferromagnets
(see e.g., Ref.~\onlinecite{Mattis}).
Another important two-particle problem is the Cooper problem, 
with two electrons in the presence of a Fermi sea
(see e.g., Ref.~\onlinecite{Baym}).
And motivated by the possibility of Cooper pair formation
in high-temperature superconductors,
there have also been a number of studies on
bound states on a two-dimensional 
lattice.\cite{Yang,Hirsch,Lin91,Petukhov,Blaer,Leung01} 
The two-electron problem 
in the plain two-dimensional repulsive Hubbard model 
is studied in Ref.~\onlinecite{Fabrizio},
and ground state energy in the large-lattice limit 
is obtained analytically.

In this section, we present a rather complete 
calculation for the two-particle problem in our model,
treating both bosons and fermions. With infinite repulsive 
interaction in our model, we are not interested in 
finding bound states. We calculate eigenenergies 
for all states for a finite-size lattice,
and our calculation is more complicated than the Hubbard 
model\cite{Fabrizio} case because of nearest-neighbor 
(in place of on-site) interaction. 
Where the Green function in the Hubbard case was a 
scalar object, in our case it is replaced by a $4\times 4$ matrix, 
corresponding to the four nearest neighbor sites where the
potential acts.  This Green function
study of the two-particle problem is closely related 
to the treatment of the two-electron problem in the
Hubbard model\cite{Fabrizio} and that in 
an extended Hubbard model.\cite{Blaer} We will show
the use of lattice symmetry and recursion relations
to simplify the problem with nearest-neighbor interactions.

\subsection{Preliminary}

In this two-particle calculation, we will work in momentum
space, and we will start with a Hamiltonian more general than 
Eq.~(\ref{eq-Ham}),\cite{Blaer}
	\begin{eqnarray}
\label{eq-Hamg}
&&H=T+U,\\
\label{eq-tg}
&&T=\sum_{\rr_1\rr_2} t(\rr_2-\rr_1) c_{\rr_1}^\dagger c_{\rr_2},\\
\label{eq-Ug}
&&U=\sum_{\rr_1\rr_2} V(\rr_2-\rr_1) c_{\rr_1}^\dagger c_{\rr_1}
c_{\rr_2}^\dagger c_{\rr_2}.
	\end{eqnarray}
Here we have allowed hopping and interaction between any two
lattice sites, but we require that both
depend only on the separation between the
two vectors and both have inversion symmetry. That is 
$t(\rr_1,\rr_2)=t(\rr_2-\rr_1)$, $t(-\rr)=t(\rr)$,
$V(\rr_1,\rr_2)=V(\rr_2-\rr_1)$, and $V(-\rr)=V(\rr)$.
In momentum space, Eqs.~(\ref{eq-tg}) and (\ref{eq-Ug}) become,
	\begin{eqnarray}
\label{eq-Tq1}
&&T=\sum_\pp \E(\pp) c_\pp^\dagger c_\pp,\\
\label{eq-Uq}
&&U=\frac{1}{2N}\sum_{\pp\pp'\kk}V(\kk)
c_\pp^\dagger c_{\pp'}^\dagger c_{\pp'+\kk} c_{\pp-\kk},
	\end{eqnarray}
where
	\begin{eqnarray}
\label{eq-eg}
&&\E(\pp)=\sum_\rr t(\rr)e^{i\pp\rr},\\
\label{eq-Vg}
&&V(\kk)=\sum_\rr V(\rr)e^{i\kk\rr},
	\end{eqnarray}
with $\E(-\pp)=\E(\pp)$ and $V(-\kk)=V(\kk)$.
Eqs.~(\ref{eq-tg}) and (\ref{eq-Ug}) reduce to our 
nearest-neighbor Hamiltonian Eq.~(\ref{eq-Ham}) if we take,
	\begin{eqnarray}
\label{eq-t}
&&t(\rr)=\cases{-t, & $\rr=(\pm 1,0)(0,\pm 1)$,\cr
0, & otherwise,}\\
\label{eq-V}
&&V(\rr)=\cases{V, & $\rr=(\pm 1,0)(0,\pm 1)$,\cr
0, & otherwise,}
	\end{eqnarray}
where we have taken the lattice constant to be unity,\cite{FN-notation}
and the nearest-neighbor vectors will be called
        \begin{equation}
        \label{eq-Rlist}
        {\RR_1}= (1,0), \RR_2=(-1,0), \RR_3=(0,1), \RR_4=(0,-1). 
        \end{equation}
Then we have,
	\begin{eqnarray}
\label{eq-ek}
&&\E(\pp)=-2t(\cos p_x+\cos p_y),\\
\label{eq-Vk}
&&V(\kk)=2V(\cos k_x+\cos k_y).
	\end{eqnarray}
Note that the structure of later equations depends sensitively on
having four sites in Eq.~(\ref{eq-V}) 
where $V(\rr) \neq 0$m, but does not depend
much on the form of Eq.~(\ref{eq-t}) 
and the resulting dispersion Eq.~(\ref{eq-ek}).

Using momentum conservation of Eq.~(\ref{eq-Hamg}),
the two-particle wave function that we will use is,
	\begin{equation}
|\psi\rangle=\sum_{\qq} g(\qq) |\qq,\PP-\qq\rangle,
	\label{eq-psig}
	\end{equation}
where the sum is over the whole Brillouin zone,
and the coefficient $g(\qq)$ satisfies,
	\begin{equation}
g(\PP-\qq)=\sbf g(\qq),
	\label{eq-g}
	\end{equation}
where $\sbf=1$ for bosons and $-1$ for fermions.

\subsection{Green function equations}

Applying the more general form of the Hamiltonian
operator Eqs~(\ref{eq-Tq1}) and (\ref{eq-Uq}) to the state 
Eq.~(\ref{eq-psig}), the Schrodinger equation 
$(E-T)|\psi\rangle=U|\psi\rangle$ becomes
	\begin{equation}
	\label{eq-sch}
(E-\E(\qq)-\E(\PP-\qq))g(\qq)=
\frac{1}{N}\sum_\kk V(\qq-\kk)g(\kk).
	\end{equation}
Eq.~(\ref{eq-sch}) is a
matrix equation ${\cal A}g=Eg$ where 
${\cal A}_{\qq\kk}=(\E(\qq)+\E(\PP-\qq))\delta_{\qq\kk}+V(\qq-\kk)/N$.
If $V$ is not infinity, this $N\times N$ matrix ${\cal A}$ 
can be diagonalized,
and $E$ and $g(\qq)$ are respectively the eigenvalue
and eigenvector. To deal with $V=+\infty$, we need
some further manipulations.

We consider the case when $E\ne\E(\qq)+\E(\PP-\qq)$, for any
$\qq$, which is to say, the energy $E$ is not the energy of
a noninteracting pair. The (lattice)
Fourier transform of the coefficients $g(\qq)$ is
	\begin{equation}
	\label{eq-ft}
	\gt(\rr)=\sum_\qq e^{-i\qq\cdot\rr} g(\qq);
	\end{equation}
this is just the real-space wavefunction in terms of the relative coordinate $\rr$.
Define the lattice Green function,
	\begin{equation}
	\label{eq-green}
	G(E,\PP;\rr,\rr')=\frac{1}{N}\sum_\qq
	\frac{e^{i\qq\cdot(\rr'-\rr)}}{E-\E(\qq)-\E(\PP-\qq)},
	\end{equation}
then after dividing the first factor from both sides of 
our Schrodinger equation, Eq.~(\ref{eq-sch}), and
Fourier transforming,  we obtain
	\begin{equation}
	\label{eq-greeneqg}
	\gt(\rr)=\sum_{\rr'}G(E,\PP;\rr,\rr')V(\rr')\gt(\rr').
	\end{equation}

In the following we return to the nearest-neighbor
potential $V(\rr)$ in Eq.~(\ref{eq-V}). 
The Green function sum in 
Eq.~(\ref{eq-greeneqg}) then has only four terms,
	\begin{equation}
	\label{eq-greeneq}
	\gt(\rr)=\sum_{j}G(E,\PP;\rr,\RR_j)(V\gt(\RR_j)),
	\end{equation}
summed over the separations in Eq.~(\ref{eq-Rlist}). 
If we also restrict $\rr$ to the four nearest-neighbor 
vectors,\cite{FN-Blaer} then Eq.~(\ref{eq-greeneq}) becomes,
	\begin{equation}
	\label{eq-gtR}
	\gt(\RR_i)=\sum_{j}G(E,\PP;\RR_i,\RR_j)(V\gt(\RR_j)).
	\end{equation}
If we define the $4\times 4$ matrix,
	\begin{equation}
	\GR_{ij}(E,\PP)=G(E,\PP;\RR_i,\RR_j),
	\label{eq-GRdef}
	\end{equation} 
and a $4\times 1$ vector $\ph_j=\gt(\RR_j)$,
then we obtain a simple matrix equation,
	\begin{equation}
        \label{eq-IminusGnull}
	(I-\GR(E,\PP)V)\ph=0.
	\end{equation}
We can also rewrite this equation as
an equation for energy using the determinant,
	\begin{equation}
	\det(I-\GR(E,\PP)V)=0.
	\end{equation}
With $V=+\infty$, we have even simpler equations 
	\begin{equation}
	\label{eq-null}
	\GR(E,\PP)(V\ph)=0,
	\end{equation}
and
	\begin{equation}
	\label{eq-det}
	\det \GR(E,\PP)=0.
	\end{equation}
Notice we write $V\phi$ to denote the limit as $V\to\infty$;
it would not do to write simply $\phi$ in Eqs.~(\ref{eq-IminusGnull}) 
and (\ref{eq-null}), since $\phi \to 0$ as $V \to \infty$
(being the amplitude of the relative wavefunction at the
forbidden separations $\{\RR_i\}$.). 

For the Hubbard model, there is only on-site interaction,
so $V(\rr)$ is nonzero only when $\rr=0$,
and the sum in Eq.~(\ref{eq-greeneqg}) has only one term.
Eq.~(\ref{eq-gtR}) is simply a scalar equation,
which, after $\gt$ cancels from both sides of the equation
and using Eq.~(\ref{eq-green}), gives,
	\begin{equation}
	1=\frac{V}{N}\sum_\qq\frac{1}{E-\E(\qq)-\E(\PP-\qq)},
	\label{eq-GHubbard}
	\end{equation}
which is exactly the result in Ref.~\onlinecite{Fabrizio}.

\subsubsection{Simplifications for rectangular boundaries}
\label{sec-eqforrec}

We specialize to the case of total momentum 
$\PP=0$ and rectangular-boundary
lattices. We have from Eqs.~(\ref{eq-green}) and (\ref{eq-GRdef}),
	\begin{equation}
\GR_{ij}(E)=\frac{1}{N}\sum_\qq
\frac{\cos(q_x(R_{jx}-R_{ix}))\cos(q_y(R_{jy}-R_{iy}))}{E-2\E(\qq)},
	\label{eq-GRij}
	\end{equation}
where the potential is nonzero on the sites $\{\RR_i\}$ given by 
Eq.~(\ref{eq-Rlist})
and in the last step we have used the symmetry properties
of the dispersion relation $\E(q_x,q_y)=\E(q_x,-q_y)=\E(-q_x,q_y)$.
Obviously Eq.~(\ref{eq-GRij}) is a function of displacements $\RR_j-\RR_i$, 
which (in view of Eq.~(\ref{eq-Rlist}) can be (0,0), (1,1), 
(2,0), or any vector related by square symmetry. 
%
It is convenient for this and later sections
to define a new notation for the Green function $\GR_{ij}$,
emphasizing its dependence on $\RR_j-\RR_i = (m,n)$, 
	\begin{equation}
	\label{eq-greenmn}
\Gamma(E,m,n)=
\frac{1}{N}\sum_\qq
\frac{\cos(m q_x)\cos(n q_y)}{E+4\cos q_x+4\cos q_y},
	\end{equation}
where the sum is over the $N$ wavevectors 
$\qq=(2\pi l_x/L_x,2\pi l_y/L_y)$ with $0\le l_x < L_x$ and 
$0\le l_y < L_y$ (for one Brillouin zone),
and we have used the expression for $\E(\qq)$
from Eq.~(\ref{eq-ek}) (and taken $t=1$).

This Green function for rectangular-boundary lattices
satisfies the following reflection properties,
	\begin{eqnarray}
	\nonumber
	&&\Gamma(E,m,n)=\Gamma(E,-m,n)\\
	&&=\Gamma(E,m,-n)=\Gamma(E,-m,-n).
	\label{eq-rectsym}
	\end{eqnarray}
And if we have a square lattice ($L_x=L_y$)
we also have
	\begin{equation}
	\Gamma(E,m,n)=\Gamma(E,n,m),
	\label{eq-squaresym}
	\end{equation}
Eq.~(\ref{eq-GRij}) can be written as,
	\begin{equation}
	\GR_{ij}(E)=\Gamma(E,R_{jx}-R_{ix},R_{jy}-R_{iy}).
	\label{eq-GRGamma}
	\end{equation}

Using the reflection properties of $\Gamma(E,m,n)$, Eq.~(\ref{eq-rectsym}),
and the definition Eq.~(\ref{eq-GRGamma}), our 
Green function matrix becomes,
	\begin{equation}
	\label{eq-Gmat}
	\GR_{ij}(E)=\left(
	\begin{array}{cccc}
	a & c & b & b\\
	c & a & b & b\\
	b & b & a & d\\
	b & b & d & a
	\end{array}
	\right),
	\end{equation}
where $a=\Gamma(E,0,0)$, $b=\Gamma(E,1,1)$, $c=\Gamma(E,2,0)$, 
and $d=\Gamma(E,0,2)$.
The eigenvalues and eigenvectors of this matrix are,
	\begin{eqnarray}
	\nonumber
	&&\lambda_{f1}=a-c, \quad (1,-1,0,0)\\
	\nonumber
	&&\lambda_{f2}=a-d, \quad (0,0,1,-1)\\
	\nonumber
	&& \lambda_{b1,b2}=a+\frac{c + d}{2}\pm 
      	\frac{\sqrt{16 b^2 + (c-d)^2}}{2},\\
	&&\quad\quad\quad (v_{1,2},v_{1,2},1,1)
	\label{eq-evals}
	\end{eqnarray}
where $v_1$ and $v_2$ are complicated functions of
$a$, $b$, $c$, and $d$.

The exact energy $E$ makes the matrix $\GR_{ij}(E)$
singular, which means that one of the eigenvalues
has to be zero. 
From Eq.~(\ref{eq-null}), the null eigenvector of $\GR$
is $V\ph=V(\gt(\RR_1),\gt(\RR_2),\gt(\RR_3),\gt(\RR_4))$, 
in terms of $\{\RR_i\}$ as in Eq.~(\ref{eq-Rlist}). 
The relative wavefunction  should be odd or even under inversion ,
depending on statistics, i.e. $\gt(-\rr)=\sbf e^{i\PP\cdot\rr}\gt(\rr)$
which follows immediately from Eqs.~(\ref{eq-ft}) and (\ref{eq-g}). 
Inversion, acting on nearest-neighbor vectors Eq.~(\ref{eq-Rlist}), 
induces $\RR_1 \leftrightarrow \RR_2$ and $\RR_3 \leftrightarrow \RR_4$;
thus with $\PP=0$, we should have $V\ph_1=-V\ph_2$
and $V\ph_3=-V\ph_4$ for fermions, and $V\ph_1=V\ph_2$
and $V\ph_3=V\ph_4$ for bosons. Inspecting the eigenvectors we obtained
in Eq.~(\ref{eq-evals}), we see that those
corresponding to $\lambda_{f1,f2}$ are antisymmetric under inversion --
corresponding to a ``p-wave-like'' (relative angular momentum 1) state
for fermions. 
Setting $\lambda_{f1}=0$, we get $a=c$,  or setting $\lambda_{f2}=0$ $a=d$, which 
respectively mean 
	\begin{eqnarray}
	\label{eq-fermioneq0}
	&&\Gamma(E,0,0)-\Gamma(E,2,0)=0, \hbox{or}\\
	&&\Gamma(E,0,0)-\Gamma(E,0,2)=0.
	\label{eq-fermioneq}
	\end{eqnarray}
Associated with the even eigenvectors are  $\lambda_{b1,b2}$ which 
are identified as boson eigenvalues.

\subsubsection{Simplifications for square boundaries}

The boson eigenvalues, Eq.~(\ref{eq-evals}), are
rather complicated for general rectangular-boundary lattices.
For a square-boundary lattice, using Eq.~(\ref{eq-squaresym}),
we get $c=d$ in the matrix Eq.~(\ref{eq-Gmat}), 
which makes the fermion eigenvalues $\lambda_{f1,F2}$ degenerate.
The boson eigenvalues
in Eq.~(\ref{eq-evals}) simplify greatly to 
$\lambda_{b1}=a+2b+c$ and $\lambda_{b2}=a-2b+c$, which means that 
the boson energy equations are,
	\begin{eqnarray}
	\label{eq-bosoneq1}
	&&\Gamma(E,0,0)+2\,\Gamma(E,1,1)+\Gamma(E,2,0)=0, \\
	\label{eq-bosoneq2}
	&&\Gamma(E,0,0)-2\,\Gamma(E,1,1)+\Gamma(E,2,0)=0.
	\end{eqnarray}
The corresponding eigenvectors simplify too, to
$(1,1,1,1)$ and $(1,1,-1,-1)$ respectively, which 
may be described as ``s-wave-like'' and ``d-wave-like'', 
i.e. relative angular momentum 0 and 2.

\subsection{Large-$L$ asymptotics for two-boson energy}
\label{sec-bosonlargeL}

Eqs.~(\ref{eq-fermioneq0}), (\ref{eq-fermioneq}),
(\ref{eq-bosoneq1}) and (\ref{eq-bosoneq2})
are much better starting points
for analytical calculations than
the original determinant equation
Eq.~(\ref{eq-det}). In the center of the 
problem is the lattice Green function
$\Gamma(E,m,n)$ defined in Eq.~(\ref{eq-greenmn}).
Many of the lattice calculations come
down to evaluating these lattice Green functions.\cite{Hirsch, 
Lin91, Petukhov, Blaer, Leung01, FN-resistor}
In this section, we derive the large-lattice
two-boson energy using the recursion 
and symmetry relations of the Green 
function $\Gamma(E,m,n)$.

The Green function $\Gamma(E,m,n)$
for general $m$ and $n$ and finite lattice are
difficult to evaluate. 
The good thing is that there are 
a number of recursion relations 
connecting the Green functions at different 
$m$ and $n$.\cite{green,Morita} 
These are trivial to derive after noting that Eq.~(\ref{eq-green})
(for $\PP=0$) can be written 
        \begin{equation}
	[E -(4+\Delta_\rr^2)-(4+\Delta_\rr'^2)] G(E,0;\rr,\rr')=
            \delta_{\rr=0}\delta_{\rr'=0}
        \end{equation}
where $\Delta_\rr^2$ is the discrete Laplacian, 
$(\Delta_\rr^2+4) f(\rr) \equiv \sum _i f(\rr+\RR_i)$ for
any function $f(\rr)$, where the sum is
over neighbor vectors Eq.~(\ref{eq-Rlist}). 
The two recursion relations that we will use are
	\begin{eqnarray}
	\label{eq-recur1}
	 E\, \Gamma(E,0,0)+4\, \Gamma(E,1,0) + 4\, \Gamma(E,0,1)&=&1,\\
	\nonumber
	 \Gamma(E,0,0)+2\,\Gamma(E,1,1)+\Gamma(E,2,0) && \\
	+\frac{1}{2} E\,\Gamma(E,1,0) &=&0.
	\label{eq-recur2}
	\end{eqnarray}
Using Eqs.~(\ref{eq-squaresym}), (\ref{eq-recur1}), and (\ref{eq-recur2}),
the boson equation Eq.~(\ref{eq-bosoneq1})
for square-boundary lattices simplifies to
	\begin{equation}
	\label{eq-bosoneq3}
	\Gamma(E,0,0)=\frac{1}{E},
	\end{equation}
with eigenvector $(1,1,1,1)$. 

Next we compute the leading form
of $\Gamma(E,0,0)$ for large $L$ of a square-boundary lattice.
The calculation is close to that in Ref.~\onlinecite{Fabrizio}
for the Hubbard model. We define $E=-8+\Delta E$. Because
the lowest energy of an independent particle is $\E({\bf 0})=-4$,
$\Delta E$ is the energy correction to two independent
particle energy at zero momentum. 
Then we have, from Eq.~(\ref{eq-greenmn}),
	\begin{eqnarray}
	\nonumber
	\Gamma(E,0,0)&=&\frac{1}{N}\sum_\qq\frac{1}{E+4\cos q_x+4\cos q_y},\\
	\nonumber
	&=&-\frac{1}{4N}\sum_\qq\frac{1}{2-\cos q_x-\cos q_y-\Delta E/4},\\
\nonumber
&\approx&
\frac{1}{L^2\Delta E}-\frac{1}{4\pi}\int_{2\pi/L}^{\pi}\frac{dq}{q}\\
&=&\frac{1}{L^2\Delta E}-\frac{\ln L}{4\pi}+{\rm const}.
	\label{eq-G00L}
	\end{eqnarray}
We should discuss the number of approximations we have made
to extract this leading dependence in $L$.
First except in the $\qq=0$ term we have ignored the
$\Delta E$ term, assuming it is small as compared
to $\qq^2$ (with $\qq\ne 0$). This is justified as
we only want the leading term in the large-$L$ limit.
Using an integral for a lattice sum is another
approximation. We choose the lower limit of integration
to be $2\pi /L$ corresponding to the first 
wavevectors after $(0,0)$ is taken out of the sum.
We also used the quadratic approximation for the  
energy dispersion $\E(\qq)$ appearing in the denominator.

Using the boson energy equation Eq.~(\ref{eq-bosoneq3})
and the large-$L$ limit of the Green function
Eq.~(\ref{eq-G00L}), we get,
	\begin{equation}
	\frac{1}{-8+\Delta E}\approx 
	\frac{1}{L^2\Delta E}-\frac{\ln L}{4\pi}+{\rm const}.
	\end{equation}
In the large-$L$ limit, $\Delta E\rightarrow 0$ (as
it is the interaction correction to the noninteracting energy),
so we get, to the leading order of $L$,
	\begin{equation}
	\Delta E=\frac{4\pi}{L^2 \ln L}.
	\label{eq-del}
	\end{equation}
We will check Eq.~(\ref{eq-del}) in Sec.~\ref{sec-afewparticles}.

\subsection{Large-$L$ asymptotics for few-particle energy}
\label{sec-afewparticles}

The procedure used in Sec.~\ref{sec-bosonlargeL}
for two bosons can also be 
applied to problems with a few particles.
For a few particles on a large lattice with short-range
(here nearest-neighbor) interaction, two-particle
interaction is the main contribution to energy.
We write for two particles,
	\begin{equation}
	E(2,L)=E_0(2,L)+\Delta E(L).
	\end{equation}
Here in this section we use
the notation $E(M,L)$ and $E_0(M,L)$ to 
denote the $M$-particle 
exact and noninteracting ground state energies
respectively and emphasize the dependence of $\Delta E$
on $L$ by using $\Delta E(L)$.
It is reasonable to expect that the 
energy for $M$ particles is the noninteracting
energy plus interaction corrections from
the $M(M-1)/2$ pairs of particles.
We then have,
	\begin{equation}
	E(M,L)\approx E_0(M,L)+\frac{M(M-1)}{2}\Delta E(L).
	\label{eq-eml}
	\end{equation}
For bosons, $E_0(M,L)=-4M$,
because in the ground state, all bosons occupy
the zero-momentum state. On the other hand, for
fermions, because of Pauli exclusion, no
two fermions can occupy the same state, the
noninteracting ground state is obtained from 
filling the $M$ fermions from the lowest state ($\kk=0$)
up.

\begin{figure}[ht]
	\centering
	\includegraphics[width=\linewidth]{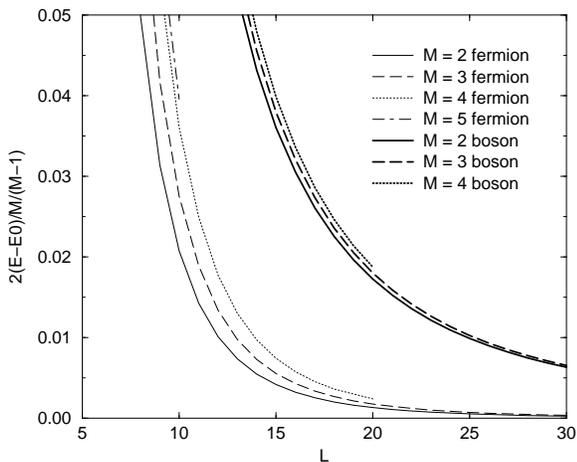}
	\caption{
Boson and fermion $2(E(M,L)-E_0(M,L))/(M(M-1))$ versus 
$L$ for $M=2,3,4,5$.
All curves appear to converge at large $L$.
The fermion (p-wave) result is much less than the boson
result (s-wave). The $M=4$ plot goes to $L=20$
and the $M=5$ plot to $L=10$. The boson
$M=5$ curve is too high to be included in this plot.
}
	\label{fig-BFlargeL}
	\end{figure}

Eq.~(\ref{eq-eml}) implies that 
plotting $2(E(M,L)-E_0(M,L))/(M(M-1))$ versus $L$ for different
$M$ should all asymptotically at 
large $L$ approach $\Delta E(L)$.
In Fig.~\ref{fig-BFlargeL}, we do such plots,
for bosons and fermions with $M=2,3,4,5$.
The fermion results, from p-wave scattering (as
our spinless fermion wave function has to be antisymmetric),
are much smaller than the boson results (bold curves) 
from s-wave scattering.

\begin{figure}[ht]
	\centering
	\includegraphics[width=\linewidth]{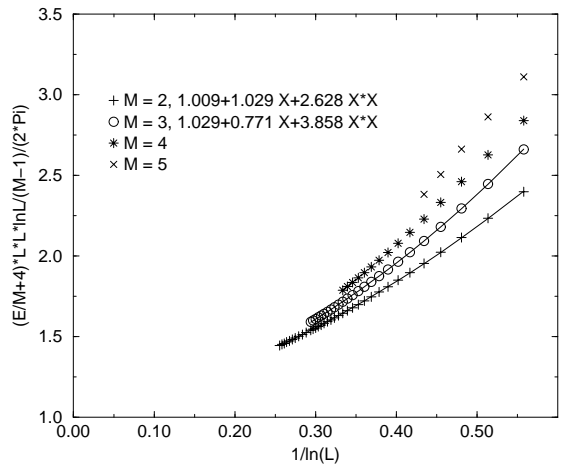}
	\caption{
Boson $(E(M,L)+4M)L^2\ln L/(2\pi M(M-1))$ versus $1/\ln L$ for $M=2,3,4,5$.
Quadratic polynomial fitting is done for $M=2$ and $M=3$.
The fitted constant coefficients are approximately one,
and the other coefficients from $M=2$ and $M=3$ are
comparable.
}
	\label{fig-bosonlargeL2}
	\end{figure}

Note that in our calculation for $\Gamma(E,0,0)$ Eq.~(\ref{eq-G00L}),
we have neglected the contribution of $\Delta E$ 
in the denominator except for the first term ($\qq=0$).
Now with the leading form of $\Delta E$ Eq.~(\ref{eq-del}),
we can obviously plug $E\approx -8+\Delta E$ into Eq.~(\ref{eq-G00L})
to get the form of the next term,
	\begin{equation}
	\label{eq-largeLfit}
\Delta E=\frac{4\pi}{L^2\ln L}\left(
A+\frac{B}{\ln L}+\frac{C}{(\ln L)^2}
\right).
	\label{eq-deltaE}
	\end{equation}
Using Eqs.~(\ref{eq-eml}) and (\ref{eq-deltaE}), we get, for 
a few bosons ($E_0(M,L)=-4M$),
	\begin{equation}
	\frac{(E(M,L)+4M)L^2 \ln L}{2\pi M(M-1)}
=A+B\frac{1}{\ln L}+C\left(\frac{1}{\ln L}\right)^2.
	\label{eq-lnL}
	\end{equation}
In Fig.~\ref{fig-bosonlargeL2}, we plot 
$(E(M,L)+4M)L^2\ln L/(2\pi M(M-1))$ versus $1/\ln L$ for $M=2,3,4,5$,
using the boson data in Fig.~\ref{fig-BFlargeL}.
Quadratic polynomial fitting is done for $M=2,3$, where we
have more data than $M=4,5$. The coefficient
$A\approx 1$ for both fits, implying, from Eq.~(\ref{eq-deltaE}), 
the leading-order term in $\Delta E(L)$
is indeed $4\pi/(L^2\ln L)$. $B$ and $C$ from two fits are
also comparable.

To summarize, from Eqs.~(\ref{eq-eml}) and (\ref{eq-deltaE}) and fitting
in Fig.~\ref{fig-bosonlargeL2}, we find that 
in our model the energy of a small number
$M$ of bosons on a large $L\times L$ lattice
is to the leading order of $L$,
	\begin{equation}
	E(M,L)\approx-4M+\frac{M(M-1)}{2}\frac{4\pi}{L^2\ln L}.
	\label{eq-fewbosons}
	\end{equation}

For two {\it fermions} on a large $L\times L$ lattice,
the noninteracting energy -- the lead term in 
Eq.~(\ref{eq-fewbosons}) -- is obviously lower for $\PP=(0,1)$ than 
for $\PP=(0,0)$.
We have not worked out the asymptotic behavior for $\PP\ne(0,0)$.

\section{The Two-Particle T-Matrix}
\label{sec-tmat}

In Sec.~\ref{sec-bosonlargeL} and
Sec.~\ref{sec-afewparticles}, we studied
the ground state energy of a few particles on a large lattice,
and we showed that the energy of $M$ particles
can be approximated by summing the energy of 
the $M(M-1)/2$ pairs. 
In this section, we reformulate the equations for two
particles and derive a scattering matrix, the t-matrix.
The t-matrix gives us equations of the form Eqs.~(\ref{eq-perturb})
and (\ref{eq-perturbmany}) which are more precise statements
of the ideas presented in Sec.~\ref{sec-bosonlargeL} and
Sec.~\ref{sec-afewparticles}. They apply to small
lattices and to excited states.

\subsection{Setup and symmetry}
\label{sec-setup}

To have an equation in the form of Eq.~(\ref{eq-perturb}),
we start with any pair of momentum vectors $\qq_1$ and $\qq_2$
and write noninteracting energy of the pair
$E_0=\E(\qq_1)+\E(\qq_2)$ and total momentum
$\PP=\qq_1+\qq_2$. Because our Hamiltonian,
Eqs.~(\ref{eq-Tq1}) and (\ref{eq-Uq}),
conserves total momentum, we can restrict
our basis states to $|\qq,\PP-\qq\rangle$.
It is tempting to take $|\qq_1,\PP-\qq_1\rangle$
and $|\qq_2,\PP-\qq_2\rangle$ as our nonperturbed
states, but there can be other two-particle states with the 
same total momentum $\PP$ and energy $E_0$.

In fact, using our energy dispersion function Eq.~(\ref{eq-ek}),
if we write $\qq_1=(q_{1x},q_{1y})$ and 
$\qq_2=(q_{2x},q_{2y})$, and define
$\qq_3=(q_{1x},q_{2y})$ and $\qq_4=(q_{2x},q_{1y})$
then we have, $\qq_1+\qq_2=\qq_3+\qq_4$
and $\E(\qq_1)+\E(\qq_2)=\E(\qq_3)+\E(\qq_4)$. 
We call this fact, that component exchanges 
in the pair $\qq_1$ and $\qq_2$ gives a pair
$\qq_3$ and $\qq_4$ that have the same total 
momentum and energy, the {\it pair component exchange 
symmetry} of our energy dispersion function
$\E(\qq)$. This symmetry is
is due to the fact that our $\E(\qq)$ is separable
into a $x$ part and a $y$ part (i.e., 
$\E(\qq)=\E_x(q_x)+\E_y(q_y)$ where 
$\E_x(q)=-2t\cos q=\E_y(q)$ in our model).\cite{FN-component}

%
%

The pair component exchange symmetry says that if
$q_{1x}\ne q_{2x}$ and $q_{1y}\ne q_{2y}$,
then the state $|\qq_3,\qq_4\rangle$, 
with $\qq_3$ and $\qq_4$ defined above using
component exchange, has the same total momentum
and energy as $|\qq_1,\qq_2\rangle$.
The degenerate perturbation theory
requires $|\qq_3,\qq_4\rangle$
should be included in the set 
of nonperturbed states with $|\qq_1,\qq_2\rangle$.

With a noninteracting
two-particle energy $E_0$ and total momentum $\PP$,
we divide the $N$ wavevectors into
two disjoint sets,
	\begin{equation}
	\QA=\{\qq\,|\,\E(\qq)+\E(\PP-\qq)=E_0\}, 
	\QB=\{\qq\,|\,\qq\not\in \QA\}.
	\label{eq-Q12}
	\end{equation}
Note that if $\qq\in \QA$ then $\PP-\qq\in \QA$. 
Denote $\NA$ the number of elements in $\QA$ and $\NB=N-\NA$ 
the number of elements in $\QB$.
With this separation of $\qq$, our eigenstate Eq.~(\ref{eq-psig})
becomes
	\begin{equation}
	\label{eq-psi2}
|\psi\rangle=\sum_{\qq\in \QA} g(\qq)|\qq,\PP-\qq\rangle
+\sum_{\qq\in \QB} g(\qq)|\qq,\PP-\qq\rangle,
	\end{equation}
where the first sum contains all states
whose noninteracting energy is degenerate.
Using the idea of degenerate perturbation
theory, we expect to be able to find a secular matrix 
$\Tm$, $\NA\times \NA$, for the degenerate states in 
$\QA$ only, and $\Tm$ will eventually be 
our momentum space t-matrix, which we will derive now.

Note that using Eq.~(\ref{eq-g}),
the number of independent
states in the first sum of Eq.~(\ref{eq-psi2})
is less than $\NA$. We include both $|\qq,\PP-\qq\rangle$
and $|\PP-\qq,\qq\rangle$ in our calculation
because we are considering boson and fermion 
problems at the same time: the symmetric solution 
$g(\qq)=g(\PP-\qq)$ is a boson solution
and the antisymmetric solution 
$g(\qq)=-g(\PP-\qq)$ is a fermion solution
(see Eq.~(\ref{eq-g})).

\subsection{Derivation of the t-matrix}
\label{sec-derivation}

Our purpose is to derive a set of closed equations
for $g(\qq)$, the coefficent in our two-particle
state Eq.~(\ref{eq-psi2}), with $\qq\in \QA$.

The Schrodinger equation for
the two-particle state $|\psi\rangle$, Eq.~(\ref{eq-sch}),
can now be written as,
	\begin{equation}
	\label{eq-schr}
	(E-\E(\qq)-\E(\PP-\qq))g(\qq)=
	\frac{1}{N}\sum_{\rr'} e^{i\qq\rr'}V(\rr')\gt(\rr'),
	\end{equation}
where $\gt(\rr)$ is the Fourier transform of $g(\qq)$
as defined in Eq.~(\ref{eq-ft}).

For $\qq\in \QB$, if we assume that 
$E\ne\E(\qq)+\E(\PP-\qq)$, Eq.~(\ref{eq-schr}) becomes
	\begin{equation}
	\label{eq-G2}
\gtA(\rr)=\gt(\rr)-\sum_{\rr'}\GB(E,\PP;\rr,\rr')V(\rr')\gt(\rr'),
	\end{equation}
where we have defined a Green function for the set $\QB$,
	\begin{equation}
	\label{eq-green2}
	\GB(E,\PP;\rr,\rr')=\frac{1}{N}\sum_{\qq\in \QB}
	\frac{e^{i\qq\cdot(\rr'-\rr)}}{E-\E(\qq)-\E(\PP-\qq)},
	\end{equation}
and a Fourier transform with vectors in $\QA$,
	\begin{equation}
	\label{eq-ft1}
	\gtA(\rr)=\sum_{\qq\in \QA} e^{-i\qq\cdot\rr} g(\qq).
	\end{equation}

By restricting to the nearest-neighbor
repulsion potential Eq.~(\ref{eq-V}), Eq.~(\ref{eq-G2}) becomes,
	\begin{equation}
	\label{eq-G2R}
\gtA(\rr)=\gt(\rr)-\sum_{j}\GB(E,\PP;\rr,\RR_j)V\gt(\RR_j)
	\end{equation}
summed over neighbor vector Eq.~(\ref{eq-Rlist}). 
Now restricting $\rr=\RR_i$ in Eq.~(\ref{eq-G2R}),
we get a set of four equations,
	\begin{equation}
	\label{eq-4eqs}
\phA_i=\ph_i-\sum_j \GRB_{ij}(E,\PP) (V\ph_j),
	\end{equation}
where we have written 
	\begin{equation}
\GRB_{ij}(E,\PP)=\GB(E,\PP;\RR_i,\RR_j)
	\label{eq-GRIIdef}
	\end{equation}
and $\ph_i=\gt(\RR_i)$ and $\phA_i=\gtA(\RR_i)$.
Eq.~(\ref{eq-4eqs}) is a matrix equation,
	\begin{equation}
	\label{eq-matrixeq}
\phA=(I-\GRB(E,\PP)V)\ph,
	\end{equation}
where $\GRB$ is $4\times 4$, $\ph$ and $\phA$
are $4\times 1$, and $V$ is a scalar (strength of potential).
And we can invert the matrix to get,
	\begin{equation}
	\label{eq-invert}
\ph=\left(I-\GRB(E,\PP)V\right)^{-1}\phA.
	\end{equation}
This is a key result, as we have expressed
the desired function $\gt$, a Fourier transform of $g(\qq)$ 
including all $\qq$, in terms of 
$\gtA$ which includes only $\qq\in \QA$; the information
about other $\qq\in \QB$ was packaged
into the Green function $\GRB(E,\PP)$.

Now we go back to Eq.~(\ref{eq-schr}), restrict
the summation to $\RR_i$, and substitute in
$V\gt_i$ from Eq.~(\ref{eq-invert}), and we get,
	\begin{equation}
	\label{eq-Gqeq}
(E-\E(\qq)-\E(\PP-\qq))g(\qq)
=\sum_{\qq'\in \QA}T(E,\PP;\qq,\qq')g(\qq').
	\end{equation}
where in the last step we have used the Fourier
transform of $\gtA(\RR_j)$ Eq.~(\ref{eq-ft1})
and defined,
	\begin{equation}
	\label{eq-Gq}
T(E,\PP;\qq,\qq')=
\frac{1}{N}\sum_{ij}e^{i\qq\RR_i}e^{-i\qq'\RR_j}
\left(V(I-\GRB(E)V)^{-1}\right)_{ij}.
	\end{equation}
If we restrict $\qq\in \QA$ in Eq.~(\ref{eq-Gqeq}), then we have,
	\begin{equation}
	\label{eq-Tqv}
(E-E_0)g(\qq)
=\sum_{\qq'\in \QA}T(E,\PP;\qq,\qq')g(\qq'),
	\end{equation}
which means,
	\begin{equation}
	\label{eq-Tq}
E=E_0+{\rm Eigenvalue}(\Tm(E)),
	\end{equation}
where we have written 
	\begin{equation}
\Tm_{\qq,\qq'}(E)=T(E,\PP;\qq,\qq')
	\label{eq-Tm}
	\end{equation}
and left out the dependence on $\PP$.
$\Tm_{\qq,\qq'}$ is the t-matrix
in momentum space. Both $\qq$ and $\qq'$ in Eq.~(\ref{eq-Tqv})
are in $\QA$, which means that if there are $\NA$ elements in $\QA$
then the matrix $\Tm(E)$ is $\NA\times \NA$.

Eq.~(\ref{eq-Tq}) is our desired equation that shows explicitly
the interaction correction to the noninteracting energy $E_0$.
In Appendix~\ref{sec-physical}, we show the physical meaning 
of $T(E,\PP;\qq,\qq')$ in the language of diagrammatic
perturbation theory, namely it is the
sum total of all the terms with repeated scattering
of the same two particles. 
This t-matrix formalism for the two-particle problem
is therefore exact, and it is exactly equivalent to the 
Schrodinger equation and the Green function
formalism in Sec.~\ref{sec-green}.
The resulting equation is an implicit
equation on $E$,  of the form $E=E_0+\Delta E(E)$
of Eq.~(\ref{eq-perturb}), and we will
show in a later section that for fermions
the approximation $E\approx E_0+\Delta E(E_0)$
is often very good.

Note also that for our case $V=+\infty$,
the t-matrix expression Eq.~(\ref{eq-Gq})
becomes
\begin{equation}
T(E,\PP;\qq,\qq')=
\frac{1}{N}\sum_{ij}e^{i\qq\RR_i}e^{-i\qq'\RR_j}
\left(-\GRB(E)^{-1}\right)_{ij},
\end{equation}
where the potential $V$ cancels out, giving a finite
value. This is one of the advantages of the 
t-matrix formalism 
that it can deal with
infinite (singular) potential, 
for which straightforward perturbation theory would diverge.

The definition of $T(E,\PP;\qq,\qq')$ in Eq.~(\ref{eq-Gq}) 
is a Fourier transform of the real space quantity
$V(I-\GRB(E)V)^{-1}$.
Here $\GRB$ is $4\times 4$ because we have
nearest-neighbor interaction. When there is only
on-site interaction, as is in the usual Hubbard model case,
$\GB(E)=\GB(E,\PP,(0,0),(0,0))$, Eq.~(\ref{eq-green2}),
is a scalar. Then, we can simply use the scalar
quantity $V/(I-\GB V)$,
which is the t-matrix that has appeared 
in Kanamori,\cite{Kanamori} Mattis,\cite{Mattis}
Rudin and Mattis,\cite{RudinMattis}
and Yosida.\cite{Yosida}
Our expression, Eq.~(\ref{eq-Gq}), is more
complicated because we have nearest-neighbor
interaction (and thus the relevance of $\RR_j$).

\subsection{Symmetry considerations}
\label{sec-symmetrycon}

In Sec.~\ref{sec-eqforrec}, 
after deriving the general Green function
equation using $\GR(E)$, we specialized 
to rectangular-boundary lattices
and used lattice reflection symmetries to diagonalize 
the $4\times 4$ matrix $\GR(E)$ and 
obtained scalar equations. 
Here our t-matrix equation Eq.~(\ref{eq-Tq}) requires us
to find the eigenvalues of the t-matrix $\Tm$.
In this section, we use particle permutation
symmetry and pair component exchange symmetry
to diagonalize the $\NA\times \NA$ t-matrix
$\Tm(E)$ for a few special cases.

\subsubsection{$\NA=1$}

There is only one momentum vector in $\QA$. 
Let us write $\QA=\{\qq_1\}$ (this implies that
$\PP-\qq_1=\qq_1$). Then there is only one 
unperturbed two-particle basis state 
$|\qq_1,\qq_1\rangle$ (see Eq.~(\ref{eq-psi2})). 
This must be a boson state, and
$\Tm(E)$ is a number. We write the resulting
scalar equation as,
	\begin{equation}
E=E_0+T_1(E).
	\label{eq-N11}
	\end{equation}

\subsubsection{$\NA=2$}

Here $\QA=\{\qq_1,\qq_2\}$ with $\qq_1+\qq_2=\PP$.
The basis states are $|\qq_1,\qq_2\rangle$
and $|\qq_2,\qq_1\rangle$. The symmetric (boson)
combination is 
$(|\qq_1,\qq_2\rangle+|\qq_2,\qq_1\rangle)/\sqrt{2}$,
and the antisymmetric (fermion) combination is
$(|\qq_1,\qq_2\rangle-|\qq_2,\qq_1\rangle)/\sqrt{2}$.
These have to be the eigenvectors of $\Tm(E)$.
And that is to say that if we define
	\begin{equation}
\SSS_2=\frac{1}{\sqrt{2}}
\left(
\begin{array}{cc}
1 & 1 \\
1 & $-1$ 
\end{array}
\right),
	\end{equation}
then we have $\SSS_2=\SSS_2^t$, $\SSS_2^2=I$, and
	\begin{equation}
\SSS_2\,\Tm(E)\,\SSS_2=
\left(
\begin{array}{cc}
T_{1,1}(E) & 0 \\
0 & T_{1,-1}(E)
\end{array}
\right).
	\end{equation}
Here $T_{1,1}(E)$ and $T_{1,-1}(E)$ are scalars that
correspond to boson and fermion symmetries respectively.
And our t-matrix equation Eq.~(\ref{eq-Tq})
is reduced to two scalar equations, 
	\begin{equation}
E=E_0+T_{1,1}(E),\quad
E=E_0+T_{1,-1}(E),
	\label{eq-N12}
	\end{equation}
for bosons and fermions respectively.
Our notation for the eigenvalues of $\Tm(E)$ is always to write
$T$ with subscripts that are the coefficients (in order) of the 
$N_0$ two-particle basis vectors.

\subsubsection{$\NA=4$}
\label{sec-N14}

Here $\QA=\{\qq_1,\qq_2,\qq_3,\qq_4\}$ with 
$\qq_1+\qq_2=\qq_3+\qq_4=\PP$.
The basis states are 
$|\qq_1,\qq_2\rangle$,
$|\qq_2,\qq_1\rangle$,
$|\qq_3,\qq_4\rangle$, and
$|\qq_4,\qq_3\rangle$.
Using particle permutation symmetry, we get
two states with even symmetries appropriate for bosons, 
which generically would be
	\begin{eqnarray}
&&a(|\qq_1,\qq_2\rangle+|\qq_2,\qq_1\rangle)+b
(|\qq_3,\qq_4\rangle+|\qq_4,\qq_3\rangle),\nonumber\\
&&-b(|\qq_1,\qq_2\rangle+|\qq_2,\qq_1\rangle)+a
(|\qq_3,\qq_4\rangle+|\qq_4,\qq_3\rangle),
	\end{eqnarray}
and two odd (fermion-type) states,
	\begin{eqnarray}
\label{eq-fstate1}
&&a(|\qq_1,\qq_2\rangle-|\qq_2,\qq_1\rangle)+b
(|\qq_3,\qq_4\rangle-|\qq_4,\qq_3\rangle),\nonumber\\
&&-b (|\qq_1,\qq_2\rangle-|\qq_2,\qq_1\rangle) +a
(|\qq_3,\qq_4\rangle-|\qq_4,\qq_3\rangle)
\label{eq-fstate2}
	\end{eqnarray}
where $a$ and $b$ are arbitrary coefficients to
be determined.

Recall $\NA=4$ means the pair $(\qq_1,\qq_2)$ 
has the same total momentum and energy as $(\qq_3,\qq_4)$, 
which may happen for various reasons.
When the reason is the pair component exchange symmetry
(of Sec.~\ref{sec-setup}),
{\it i.e.} $\qq_3=(q_{1x},q_{2y})$ and $\qq_4=(q_{2x},q_{1y})$, 
then $a=b=1/2$, due to a hidden symmetry under
the permutation $1 \leftrightarrow 3, 2\leftrightarrow 4$. 
The only effect this permutation has on the 
momentum transfers $\qq_i-\qq_j$ is to change the sign of
one or both components; but the potential $V(\rr)$ is
is symmetric under reflection through either coordinate
axis, hence $V(\qq_i-\qq_j)$ is invariant under the permutation.
Since the t-matrix depends only on $V(\qq_i-\qq_j)$, it 
inherits this symmetry. 
Next, if we define
	\begin{equation}
\SSS_4=\frac{1}{2}
\left(
\begin{array}{cccc}
1 & 1 & 1 & 1\\
1 & 1 & $-1$ & $-1$\\
1 & $-1$ & 1 & $-1$ \\
1 & $-1$ & $-1$ & 1
\end{array}
\right),
	\end{equation}
then we have $\SSS_4=\SSS_4^t$, $\SSS_4^2=I$, and
$\SSS_4\,\Tm(E)\,\SSS_4$
becomes diagonal with four eigenvalues of $\Tm(E)$:
$T_{1,1,1,1}(E)$, $T_{1,1,-1,-1}(E)$,
$T_{1,-1,1,-1}(E)$, and $T_{1,-1,-1,1}(E))$.
And our t-matrix equation Eq.~(\ref{eq-Tq})
is reduced to 
	\begin{equation}
E=E_0+T_{1,1,1,1}(E),\quad
E=E_0+T_{1,1,-1,-1}(E),
	\label{eq-N14b}
	\end{equation}
for bosons and
	\begin{equation}
E=E_0+T_{1,-1,1,-1}(E),\quad
E=E_0+T_{1,-1,-1,1}(E),
	\label{eq-N14f}
	\end{equation}
for fermions.

The three cases $\NA=1,2$, and $\NA=4$ with 
pair component
exchange symmetry are three special cases
in which we know the eigenvectors of $\Tm$
and can therefore diagonalize $\Tm$ from symmetry
considerations easily. 

Different or larger values of $\NA$ are possible
when $\PP$ has a special symmetry, e.g. 
when $P_x=P_y$, $\NA=8$ generically since
$\QA$ includes pairs such as $(q_{1y},q_{1x}), 
(q_{2y},q_{2x})$. 
For these general cases, 
we return to Eq.~(\ref{eq-Tq}) and diagonalize
$\Tm$ numerically.
For example, on a $L\times L$ lattice,
the pairs (0,1)$(0,-1)$ and (1,0)$(-1,0)$
have the same total energy and momentum,
but this is not due to the pair component
exchange symmetry. In this case,
we numerically diagonalize the $4\times 4$ matrix
$\Tm(E)$, and we find that in the fermion eigenvectors,
Eqs.~(\ref{eq-fstate1}) and (\ref{eq-fstate2}),
$a\ne b$.

\subsection{Solving for energy}
\label{sec-solveforE}

The example system that we will study here
is $10\times 11$ with $\PP=(0,0)$.
The noninteracting and 
interacting energies of the system are
in Table~\ref{t-10by11}. 
It can be seen that all of the energies listed 
in Table~\ref{t-10by11} are of the three cases 
discussed in Sec.~\ref{sec-symmetrycon}:
$\NA=1$, $\NA=2$, and $\NA=4$ due to
pair component exchange symmetry.

\begin{table}[ht]
        \centering
        \caption{
The 12 low-lying noninteracting and exact 
two-particle energies of the $10\times 11$ lattice
with total momentum $\PP=(0,0)$.
$\qq_1$ and $\qq_2=\PP-\qq_1$ are the momentum
vectors. 
}
\begin{ruledtabular}
        \begin{tabular}{ccc|c|c} 
$\qq_1$ & $\qq_2$ & $\E(\qq_1)+\E(\qq_2)$ & boson & fermion \\\hline
$(0,0)$ & $(0,0)$ &  -8.0000000000	& -7.9068150537	& -7.3117803781\\
$(0,1)$ & $(0,-1)$ & -7.3650141313	& -7.2998922545	& -7.1770594424\\
$(1,0)$ & $(-1,0)$ & -7.2360679774	& -6.9713379459	& -6.4994071102\\
$(1,-1)$ & $(-1,1)$& -6.6010821088  & -6.6010821088	& -6.4700873024\\
$(1,1)$ & $(-1,-1)$& -6.6010821088  & -6.0227385416	& -5.5449437453\\
$(0,2)$ & $(0,-2)$ & -5.6616600520	& -5.4277094111	& -5.1475674826\\
$(2,0)$ & $(-2,0)$ & -5.2360679774	& -5.0769765528	& -4.8309218202\\
$(1,2)$  & $(-1,-2)$& -4.8977280295  & -4.8977280295	& -4.7226011845\\
$(1,-2)$  &  $(-1,2)$& -4.8977280295  & -4.6571944706	& -4.3808316899\\
$(2,-1)$&  $(-2,1)$& -4.6010821088  & -4.6010821088	& -4.1884725717\\
$(2,1)$ & $(-2,-1)$& -4.6010821088  & -3.5439149838	& -3.3270813673\\
$(0,3)$ & $(0,-3)$ & -3.4307406469  & -3.1234645374	& -2.8242092883
        \end{tabular}
\end{ruledtabular}
        \label{t-10by11}
\end{table}

We solve for energy $E$ in the implicit equation,
$E=E_0+T(E)$, where $T(E)$ represents 
the eigenvalues of $\Tm(E)$, e.g., $T_{1,-1}(E)$.
We plot $f(E)=E_0+T(E)$ along with a line $y=E$. 
Their intersections are the desired energies $E$.

\subsubsection{$\NA=1$ case}

\begin{figure}[ht]
\centering
\includegraphics[width=\linewidth]{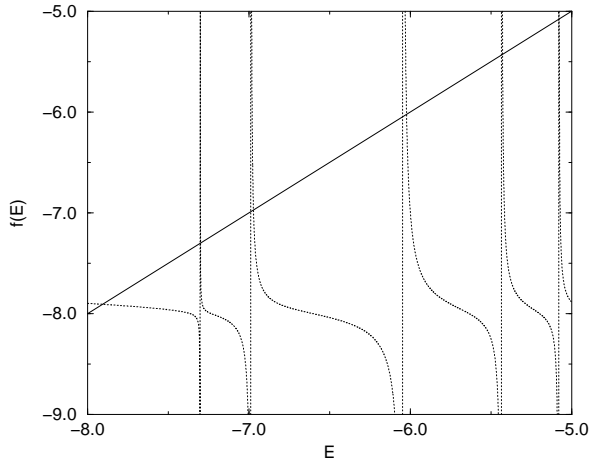}
\caption{
$f(E)=E_0+T_1(E)$
versus E for $10\times 11$ lattice with $\PP=(0,0)$ 
and $E_0=-8.0$ (i.e., $E_0=\E(0)+\E(0)$.)
The intersections with the line $y=E$ are the exact
two-particle energies.
}
	\label{fig-Tq1}
	\end{figure}

In Fig.~\ref{fig-Tq1},
we plot $f(E)$ versus $E$ for the $10\times 11$ lattice with $\PP=(0,0)$
and the noninteracting energy $E_0=-8.0=\E(0)+\E(0)$. 
Here $\QA=\{(0,0)\}$, and the nonperturbed state
is $|\qq_1=(0,0),\PP-\qq_1=(0,0)\rangle$ which can only 
be a boson state. The energy
intersections from Fig.~\ref{fig-Tq1}
are $-7.906$, $-7.299$, $-6.971$,
$-6.022$, and so on. Looking into Table~\ref{t-10by11},
we see that these are all boson energies.

In Fig.~\ref{fig-Tq1}, note also that the energy $-6.601$, which is an exact
eigenenergy from exact diagonalization, does not appear
as an intersection in Fig.~\ref{fig-Tq1}. This is a special energy,
being also a noninteracting energy. Earlier,
as mentioned at the beginning of
Sec.~\ref{sec-derivation}, we assumed that
our $E\ne\E(\qq)+\E(\PP-\qq)$ for any $\qq\in \QB$,
so this energy is excluded from our t-matrix formulation.
We will address later in Sec.~\ref{sec-4case} this kind of exact solutions 
that are also noninteracting energies.

Note that our equation $E=E_0+T(E)$ is a reformulation
of the Schrodinger equation with certain symmetry
considerations,
and it should be satisfied by all energies 
$E$ with the same symmetry. Building 
$T(E)$ from $E_0$ and $\PP$ does not automatically
give us a unique interacting energy $E$ that corresponds
to the noninteracting energy $E_0$. 
However, we can see clearly from Fig.~\ref{fig-Tq1}, 
if we perturb the exact solutions by 
a small amount $E\rightarrow E+\delta$, 
then $f(E)$ changes drastically
except for the lowest energy $E=-7.906$.
That is to say that these other energies,
for example $E=-6.971$, are
exact solutions of the equation $f(E)=E$,
but they are not stable solutions. From the
plot, only $E=-7.906$ comes close to being stable.

We can be more precise about this notion
of stability. If we have an iteration
$x_{n+1}=f(x_n)$, and $x^*$ is a fix point
(i.e., $f(x^*)=x^*$), then the iteration
is linear stable at $x^*$ if and only if
$|f'(x^*)|<1$. In our plots, we have included a
line $y=E$ with slope one, which can be used as a
stability guide. An intersection (fix point)
is linearly stable when the function 
$f(E)$ at the intersection is not 
as steep as the straight line.

\subsubsection{$\NA=2$ case}

\begin{figure}[ht]
\centering
\includegraphics[width=0.9\linewidth]{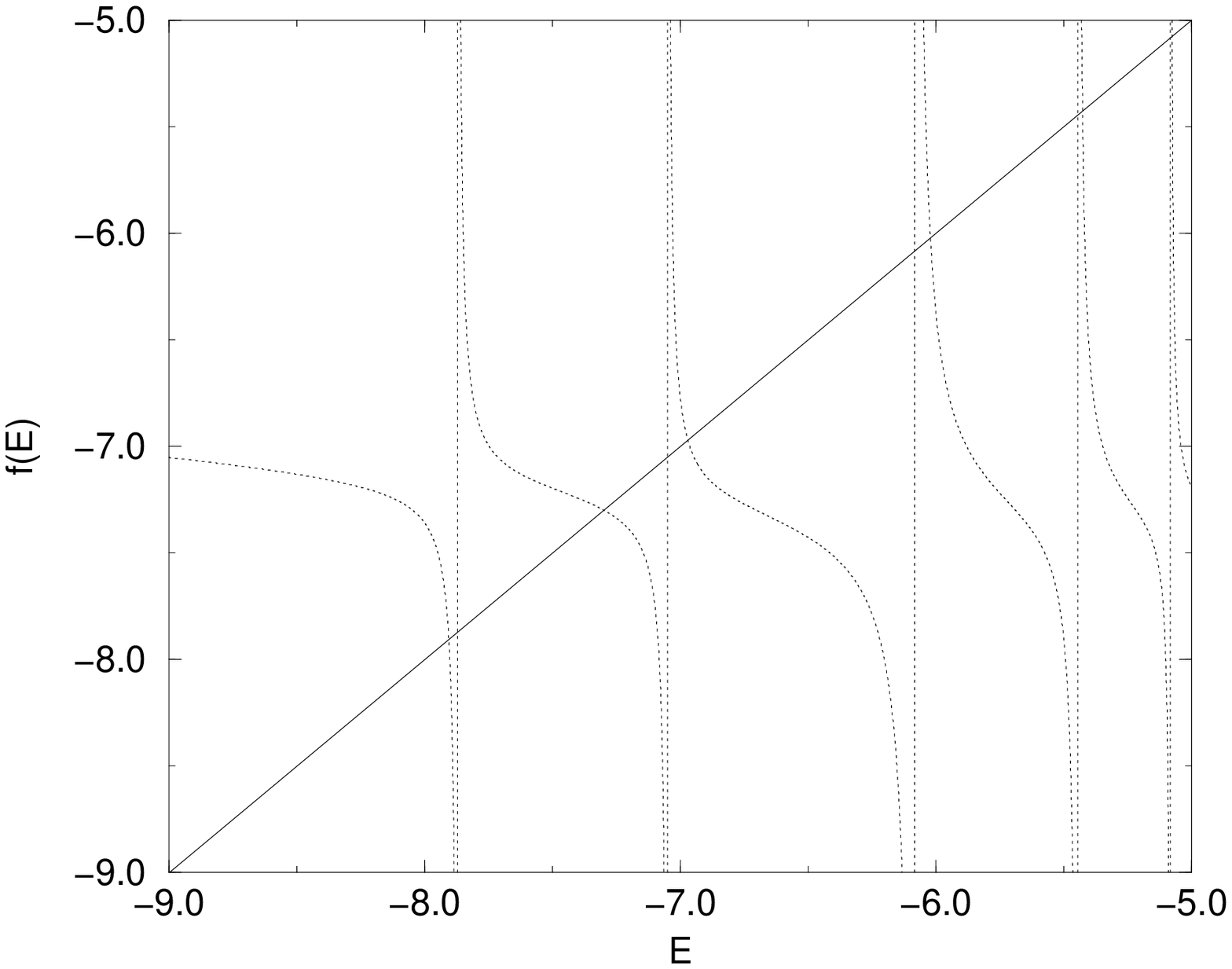}
\includegraphics[width=0.9\linewidth]{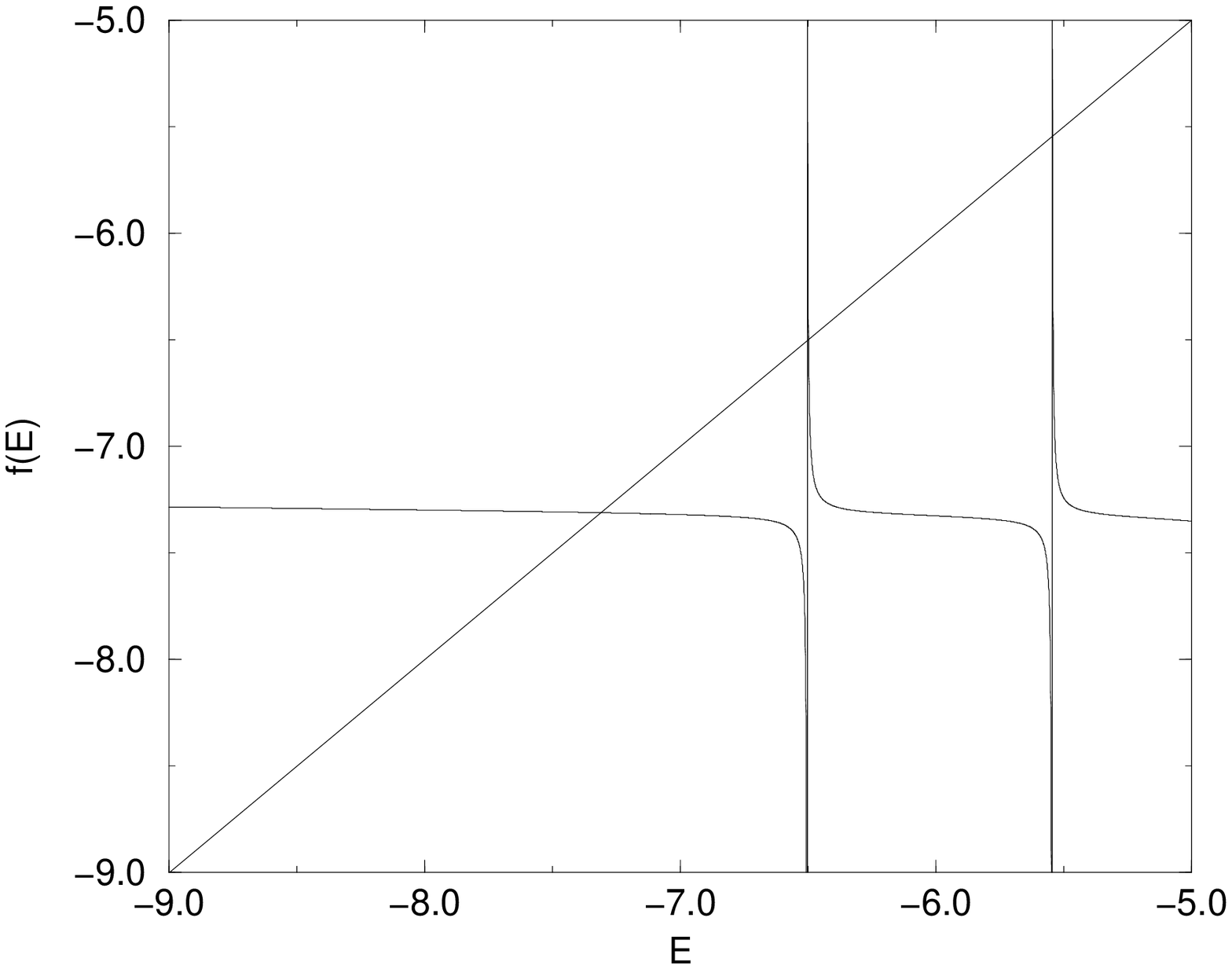}
\caption{
$f(E)$ versus E for $10\times 11$ lattice with $\PP=(0,0)$ 
and $E_0=-7.365$ (i.e., $E_0=\E(0,1)+\E(0,-1)$.)
The top graph (dotted line) is for boson $f(E)=E_0+T_{1,1}(E)$,
and the top graph (solid line) for fermion $f(E)=E_0+T_{1,-1}(E)$.
The fermion curve is essentially flat near $E=E_0$.
}
\label{fig-Tq2}
\end{figure}

In Fig.~\ref{fig-Tq2} we plot for $E_0=-7.365$ 
and $\PP=(0,0)$ with $\QA=\{(0,1),(0,-1)\}$.
The boson function $f(E)=E_0+T_{1,1}(E)$ is the dotted 
line in the top graph, and the fermion function 
$f(E)=E_0+T_{1,-1}(E)$ is the solid line in the 
bottom graph.

The intersections closest to $E_0=-7.365$ are 
$-7.299$, the first excited boson energy 
(see Table~\ref{t-10by11}), and $-7.311$, the 
lowest fermion energy.
Note that the curve on which the fermion intersection 
($-7.311$) lies is very flat. In other words for this fermion
energy $E\approx E_0+T(E_0)$, i.e., the first iteration
using the noninteracting energy gives an energy very close to 
the exact value. More precisely, we find with $E_0=-7.365014$,
$f(E_0)=E_0+T(E_0)=-7.310584$, which is very close
to $E=-7.31178$. Many t-matrix 
calculations,\cite{Kanamori,Mattis,RudinMattis,Yosida}
use the first iteration $E\approx E_0+T(E_0)$ as 
an approximation to the exact energy, and we see in this case
this approximation is very good. (We will come back
to this point later in Sec.~\ref{sec-nontoin}.)

\subsubsection{$\NA=4$ case}
\label{sec-4case}

\begin{figure}[ht]
\centering
\includegraphics[width=\linewidth]{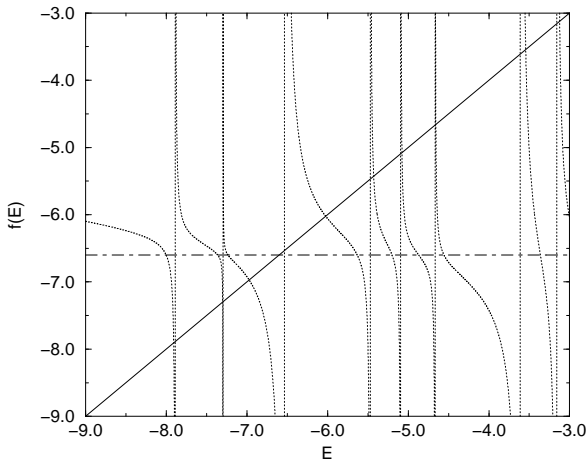}
\caption{
Boson $f(E)$ versus E for $10\times 11$ lattice with $\PP=(0,0)$ 
and $E_0=-6.601$. The dotted line is
for $T_{1,1,1,1}$ and the horizontal dot-dashed line
for $T_{1,1,-1,-1}$ (which corresponds to
a noninteracting state, see text at the end of this section). 
}
\label{fig-Tq3b}
\end{figure}

\begin{figure}[ht]
\centering
\includegraphics[width=\linewidth]{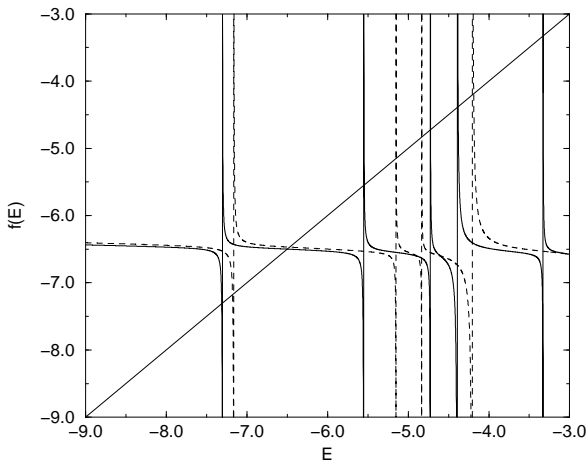}
\caption{
Fermion $f(E)$ versus E for $10\times 11$ lattice with $\PP=(0,0)$ 
and $E_0=-6.601$. The solid line is
for $T_{1,-1,1,-1}$ and the long-dashed line
for $T_{1,-1,-1,1}$. Note that closely spaced
fermion energy pairs are separated by symmetry.
}
\label{fig-Tq3f}
\end{figure}

In Figs.~\ref{fig-Tq3b} and \ref{fig-Tq3f},
we plot $f(E)$ for 
$E_0=\E(1,-1)+\E(-1,1)=\E(1,1)+\E(-1,-1)=-6.601$
and $\PP=(0,0)$.
For this $\NA=4$ case we have 
two boson functions, plotted in Fig.~\ref{fig-Tq3b}, 
$f(E)=E_0+T_{1,1,1,1}(E)$ (dotted line)
and $f(E)=E_0+T_{1,1,-1,-1}(E)$ (dot-dashed line),
and we have two fermion functions, plotted in Fig.~\ref{fig-Tq3f}, 
$f(E)=E_0+T_{1,-1,1,-1}(E)$ (solid line)
and $f(E)=E_0+T_{1,-1,-1,1}(E)$ (dashed line).
The fermion intersections closest to $E_0$
are $-6.499$ and $-6.470$. Here again the two fermion
curves are very flat.
The two boson intersections closest to $E_0$
are $-6.022$ and $-6.601$. Note that the latter is
also a noninteracting energy, and it is
the intersection of the horizontal line $E=E_0$ with 
$y=E$.

One interesting observation of the fermion
plot in Fig.~\ref{fig-Tq3f} is
that pairs of closely spaced energies
(for example $-7.311$ and $-7.177$)
lie on different symmetry curves.
We know that if we have a square lattice
(for example $10\times 10$) then the noninteracting
fermion energies come in pairs. Here, we have chosen
a $10\times 11$ lattice that is close to a square 
but does not have exact degeneracies.
We see that the resulting closely spaced pairs
are separated by symmetry considerations.

Another interesting observation from 
Fig.~\ref{fig-Tq3b} for bosons is that we have a horizontal line
that corresponds to $T_{1,1,-1,-1}(E)=0$. For this case 
the noninteracting energy is an exact energy. That is to say,
$(1,1,-1,-1)$ is a null vector of $\Tm(E)$ (see Sec.~\ref{sec-N14}),
or the eigenstate,
\begin{equation}
|\qq_1,\qq_2\rangle+|\qq_2,\qq_1\rangle-
|\qq_3,\qq_4\rangle-|\qq_4,\qq_3\rangle,
\end{equation}
with $\qq_3=(q_{1x},q_{2y})$ and $\qq_4=(q_{2x},q_{1y})$
is an exact eigenstate of the Hamiltonian.
This can be shown easily using 
the Schrodinger equation Eq.~(\ref{eq-sch}).
We have $g(\qq_1)=g(\qq_2)=1$,
$g(\qq_3)=g(\qq_4)=-1$, and $g(\qq)=0$ for all other
$\qq$, and we can easily show
$V(\qq-\qq_1)+V(\qq-\qq_2)-V(\qq-\qq_3)-V(\qq-\qq_4)=0$
(because $V(\kk)$ can be 
separated into a sum of two terms that
involve the $x$ and $y$ components separately).

Transforming to the real space, 
without worrying about normalization, we can have
\begin{eqnarray}
\nonumber
\gt(\rr)&=&\sum_\qq e^{-i\qq\cdot\rr} g(\qq)\\
&\sim&\left( e^{-iq_{1x} x} - e^{-iq_{2x} x}\right) 
\left( e^{-iq_{1y} y} - e^{-iq_{2y} y}\right),
\label{eq-bosonexact}
\end{eqnarray}
where we have used the fact mentioned above that 
$g(\qq)$ is not zero for only four $\qq$'s
which are related by pair component exchange symmetry.
It is clear from Eq.~(\ref{eq-bosonexact}) that
$\gt(0,y)=0=\gt(x,0)$, which means that
the wave function in relative position space
is ``d-wave'' like, having nodes along $x$ and $y$ axes
(thus happens to have nodes at every relative position where
the potential would be nonzero).

\subsection{Fermion: noninteracting to interacting}
\label{sec-nontoin}

In this section we use the t-matrix techniques
developed in the preceding sections of this section
to study the relationship between the noninteracting
energies and the interacting energies.
We start with the table of energies in
Table~\ref{t-10by11} for the $10\times 11$
lattice with $\PP=(0,0)$. We have asked in
the introduction to this section whether we can 
go from the noninteracting to the interacting
energies 
and now we know that we have
an equation $E=E_0+T(E)$ where $T(E)$
is the symmetry reduced scalar
t-matrix function.
 From our graphs (Fig.~\ref{fig-Tq2} and Fig.~\ref{fig-Tq3f})
we have commented that for fermions the curve of 
$T(E)$ around $E_0$ is quite flat (which
is not the case for bosons). And we mentioned
that this implies that the approximation $E\approx E_0+T(E_0)$
is close to the exact energy. Now in this section,
we study the t-matrix approach for a specific system.
We will denote $E_1=E_0+T(E_0)$, the first iteraction
result, and $E_{n+1}=E_0+T(E_n)$, the $n$th iteration
result.

\begin{figure}[ht]
\centering
\includegraphics[width=\linewidth]{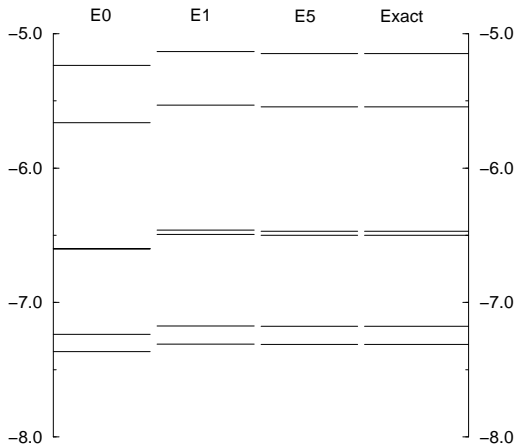}
\caption{
Two-fermion energy levels for the $10\times 11$ lattice with $\PP=(0,0)$.
 From left to right,
the lowest few noninteracting energies $E_0$, first t-matrix
iteration $E_1$, fifth t-matrix iteration $E_5$,
and the exact energy $E_{\rm exact}$ are plotted.
Note that the third noninteracting energy from the bottom
is doubly degenerate (see Table~\ref{t-10by11tmat}).
}
\label{fig-levels}
\end{figure}

In Table~\ref{t-10by11tmat} we show the
t-matrix calculation for the $10\times 11$ lattice.
We show for the lowest few states
the noninteracting energy $E_0$, the first
t-matrix iteration $E_1$, the fifth t-matrix
iteration $E_5$, and the exact energy $E_{\rm exact}$.
In Fig.~\ref{fig-levels} these energy levels are 
plotted graphically. From the table, it is 
clear that the first t-matrix iteration result $E_1$
is quite close to the exact energy,
and the fifth iteration result $E_5$ gives
a value that is practically indistinguishable
from the exact value.

\begin{table*}[ht]
        \centering
        \caption{
Fermion energies for $10\times 11$ lattice with $\PP=(0,0)$. 
$E_0=\E(\qq_1)+\E(\qq_2)$ is the noninteracting energy.
$E_n=E_0+T(E_{n-1})$ where $T(E)$ is the symmetry reduced t-matrix.
Here only fermion energies (from $T_{1,-1}$ or $T_{1,-1,1,-1}$ and $T_{1,-1,-1,1}$) are included. 
}
\begin{ruledtabular}
        \begin{tabular}{cccccc} 
$\qq_1$& $\qq_2$& $E_0$& $E_1$& $E_5$ & $E_{\rm exact}$\\\hline
(0,1)&	(0,-1)&	-7.365014&	-7.310598893&	-7.311780378&	-7.311780378 \\
(1,0)&	(-1,0)&	-7.236067&	-7.17521279&	-7.17705944&	-7.177059442 \\
(1,-1)&	(-1,1)&	-6.601082&	-6.493807907&   -6.49940706&	-6.49940711 \\
(1,1)&	(-1,-1)& -6.601082&	-6.460962404&	-6.470087137&	-6.470087302 \\
(0,2)&	(0,-2)&	-5.661660&	-5.532751985&	-5.54494225&	-5.544943745 \\
(2,0)&	(-2,0)&	-5.236067&	-5.134290466&	-5.147558003&	-5.147567483 
        \end{tabular}
\end{ruledtabular}
        \label{t-10by11tmat}
\end{table*}

\section{A Few Fermions: Shell Effect and T-Matrix}
\label{sec-tmat2}

In Sec.~\ref{sec-green}, we used lattice Green function
to study the problem of two particles (bosons and fermions),
and at the end of that section, in Sec.~\ref{sec-afewparticles},
we obtained the ground state energy of a few particles on a large lattice
by summing up the energy of each pair of particles. This section
contains a much more detailed study of the few-fermion problem: we will
consider first the fermion shell effect and then we will
study the interaction correction to energy (ground state
and excited states) for a few fermions (three, four, and five)
using the t-matrix. 

Our results -- summarized in Sec.~\ref{sec-TMerrors} --
confirm that, in the dilute limit, 
almost all of the interaction correction is accounted for by 
the {\it two-body} terms of the t-matrix approximation, 
Eq.~(\ref{eq-multiTM}). 
But (recall Eq.~(\ref{eq-perturbmany}))
that is a hallmark of a Fermi liquid picture;
{i.e.}, our numerical results 
suggest its validity at low densities. 
This is a nontrivial result, in that 
firstly, the validity of Fermi liquid theory in a finite-system
context has rarely been considered.  Standard t-matrix theory depends
on a Fermi surface which (at $T=0$) is completely sharp in momentum 
space, and every pair's t-matrix excludes scattering into the same
set of occupied states.  In a finite system, however, the allowed $\qq$ vectors
fall on a discrete grid, and since the total number of particles is finite, 
the t-matrices of different pairs see a somewhat different set of excluded states
(since they do not exclude themselves, and one particle is a non-negligible
fraction of the total). 

Secondly, and more essentially, the analytic
justifications of Fermi liquid theory exist only in the cases of spinfull fermions
(in a continuum). That case is dominated by s-wave scattering, so that 
the t-matrix approaches a constant in the limit of small momenta
(and hence in the dilute limit).  Our spinless case is rather different, 
as will be elaborated in Sec.~\ref{sec-dilute}, because the t-matrix is
dominated by the p-wave channel, which vanishes at small momenta. 
Thus the $\qq$ dependence is crucial in our case, and the numerical
agreement is less trivial than it would be for s-wave scattering. 

In this section, after an exhibition of the shell effect (Sec.~\ref{sec-shelleffect}), 
we present a general recipe for the multi-fermion t-matrix calculation. 
This is developed by the simplest cases, chosen to clarify when degeneracies
do or do not arise. 

\subsection{Fermion shell effect}
\label{sec-shelleffect}

At zero temperature,
the ground state of noninteracting fermions is formed by
filling the one-particle states one by one from the lowest
to higher energies. For our model of spinless fermions on a 
square lattice, we have the two ingredients for the shell effect:
fermionic exclusion and degeneracies of one-particle states
due to the form of our energy function and lattice symmetry.
Shell effects have been noted previously in interacting 
models;~\cite{Shell} our code, permitting 
non-rectangular boundary conditions, 
allows us to see even more cases of them

\begin{figure}[ht]
	\centering
	\includegraphics[width=0.9\linewidth]{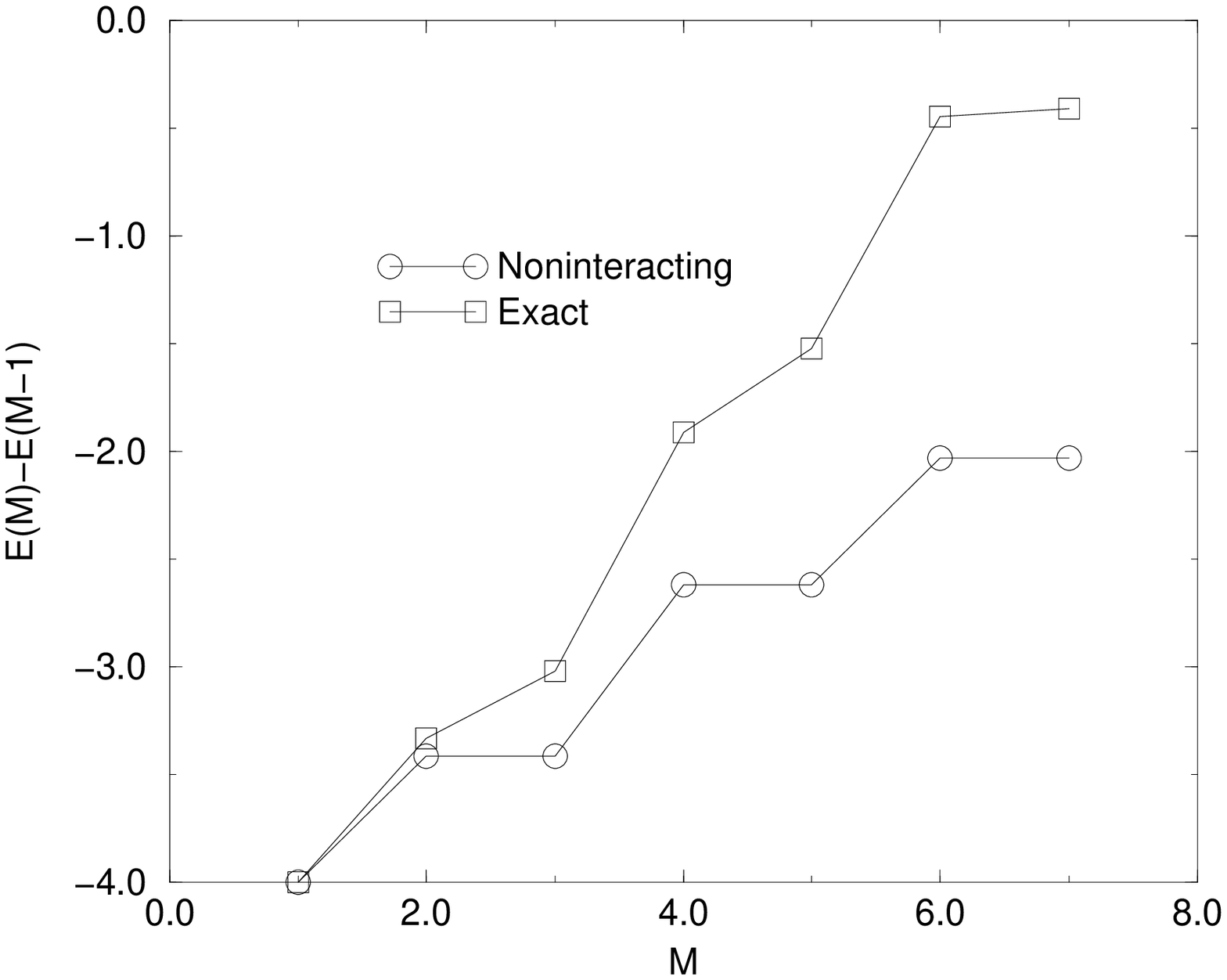}
	\includegraphics[width=0.9\linewidth]{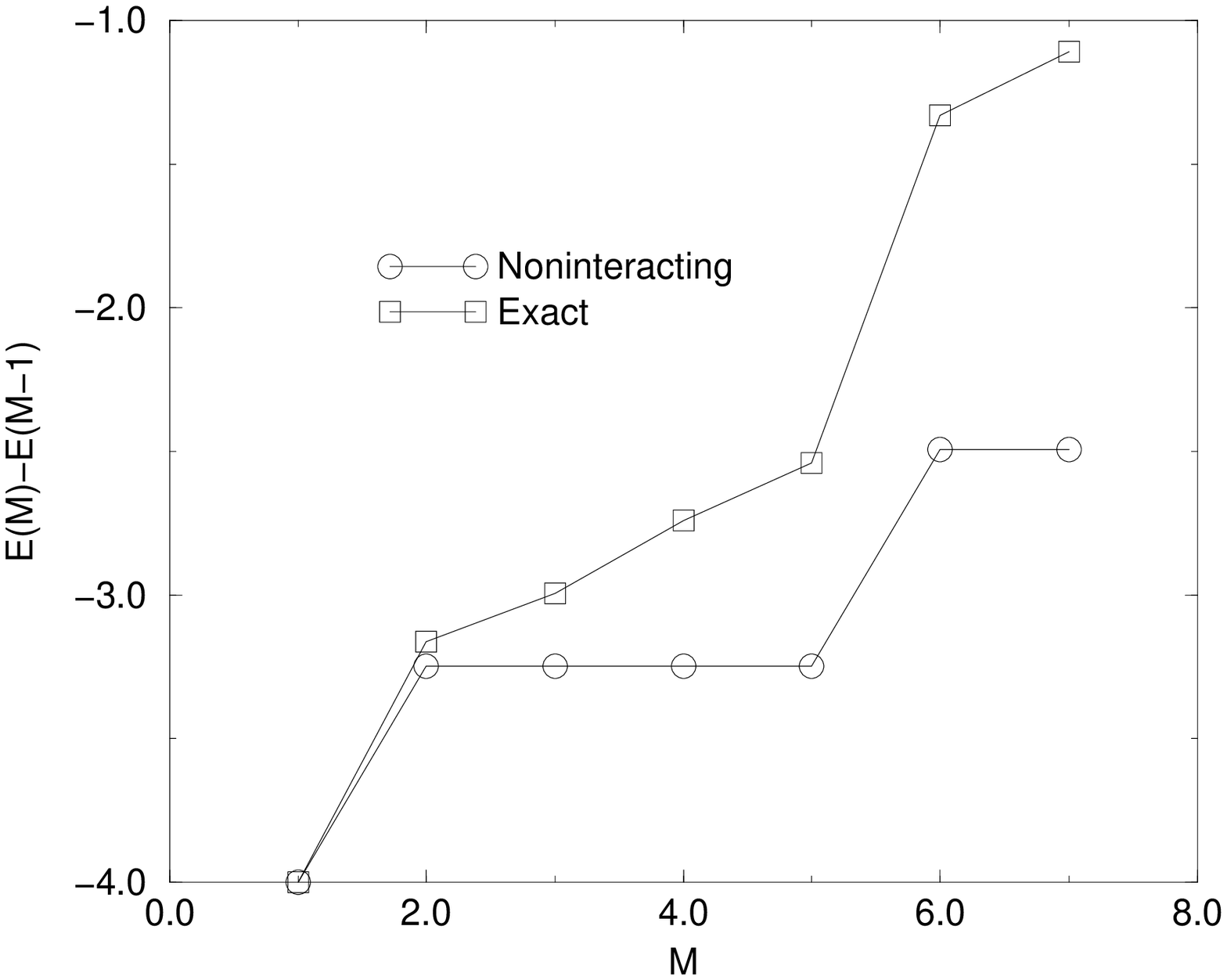}
	\caption{
	Shell effect for $5\times 8$ and $7\times 7$ 
lattices. Exact, interacting groundstate energies 
are compared with noninteracting energies for up 
to seven particles. Energy increment $E(M)-E(M-1)$
is shown. 
	}
	\label{fig-shell}
	\end{figure}

In Fig.~\ref{fig-shell} we show the exact and 
for comparison the noninteracting ground state energies
for the $5\times 8$ and $7\times 7$ lattices 
for up to seven particles.
The energy increment curve $E(M)-E(M-1)$ is plotted
and shows clearly the shell effect.

\begin{figure}[ht]
	\centering
	\includegraphics[width=0.9\linewidth]{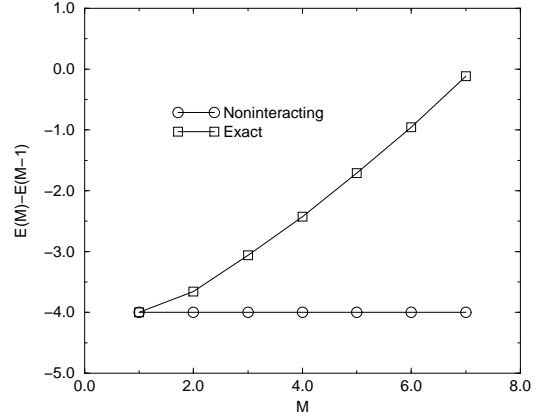}
	\caption{
Boson noninteracting and exact groundstate 
energies for the $5\times 8$ lattice with one to seven particles.
Because bosons can all be at the zero-momentum state, 
where the energy is $-4$, the noninteracting energy 
is $-4M$. The exact energy shows smooth changes when $M$ increases.
There is no shell effect.
	}
	\label{fig-boson58}
	\end{figure}

The filled shells for the $5\times 8$ lattice
are at $M=3$ (with momentum vectors $(0,0)(0,\pm 1)$ 
occupied) and $M=5$ (with $(0,0)(0,\pm 1)(\pm 1,0)$
occupied). On the other hand, $M=3$ is not a 
filled shell of the $7\times 7$ lattice.
For comparison, we show the boson energy plot
for the $5\times 8$ lattice in Fig.~\ref{fig-boson58}.
Because bosons can all be at the zero-momentum state, where energy 
is $-4$, the total noninteracting energy is $-4M$.
The exact energy curve shows smooth changes when $M$ increases.
There is no shell effect.

\subsection{General multi-fermion theory}
\label{sec-multiTM}

The key notion for generalizing our two-fermion approach 
to $M$ fermions is that the set $\QA$ now consists of every $M$-tuple $\alpha$
of wavevectors
that gives the same total momentum and noninteracting energy.
This defines a reduced Hilbert space, with the corresponding
basis states $|\phi_\alpha\rangle$. 
We can construct an approximate, effective Hamiltonian  $H_0+H_{\rm tm}$
acting within $\QA$-space, where $H_{\rm tm}$ is a sum of {\it pairwise} 
t-matrix terms, each of which changes just two fermion occupancies:
  \begin{equation}
  \label{eq-multiTM}
    H_{\rm tm} = {\sum}' _{\alpha \beta}  \T_{\alpha \beta}
  \end{equation}
The notation ${\sum}'_{\alpha\beta}$ means the sum only includes
the pair $(\alpha,\beta)$ when $|\phi_\alpha\rangle$ differs
from $|\phi_\beta\rangle$ by a change of two fermions.

Thus, each term in Eq.~(\ref{eq-multiTM})
is associated with a particular fermion pair $(\qq_i,\qq_j)$. 
Each such pairwise t-matrix can be viewed as a sum of all possible 
repeated scatterings of those two
fermions through intermediate states, except that intermediate states 
which are already included in $\QA$ are excluded.
(The most important omission of this approximation 
would be the processes in which three or more 
fermions are permutated before the system returns to the $\QA$ Hilbert space.)
Each term is a two-fermion t-matrix calculated according to
the recipe of Secs.~\ref{sec-derivation} and \ref{sec-solveforE} --
thus each term has its own two-fermion wavevector set $\QA^{i,j}$
and complementary set $\QB^{i,j}$, 
as defined in Eq.~(\ref{eq-Q12}).
The only change in the recipe is to augment the 
set $\QB^{i,j}$  of wavevectors forbidden in the intermediate
scatterings of the two fermions,  
since they cannot scatter into states already occupied by
the other fermions in states $\alpha$ and $\beta$.
(See Eq.~(\ref{eq-QBthree}) for an example.)


The t-matrix treatment is a form of perturbation expansion, for which the
small parameter is obviously not 
$V$ (which is large) but 
instead $1/L^2$, as is evident from Eq.~(\ref{eq-del}).
That is, as the lattice size is increased
(with a fixed set of fermions), 
the approximation captures a larger and larger fraction
of the difference $E_{\rm exact}-E_0$.

\subsection{A three-fermion t-matrix calculation}
\label{sec-threefermion}

\begin{figure}[ht]
        \centering
	\includegraphics[width=0.9\linewidth]{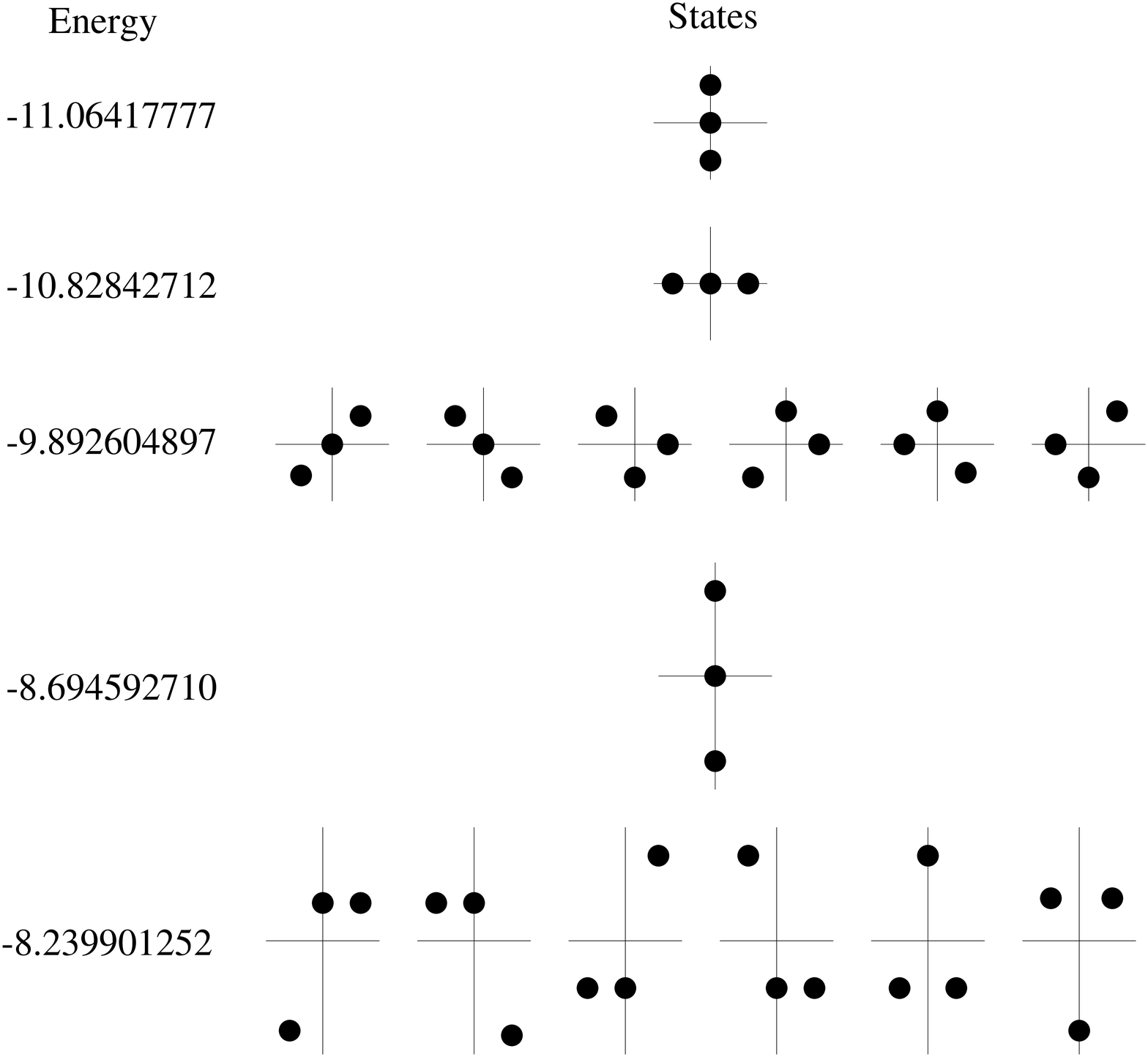}
        \caption{
Lowest five noninteracting energy levels for the 
$8\times 9$ lattice with $M=3$ fermions
and total momentum $\PP=(0,0)$.
States in momentum space are drawn.
	}
        \label{fig-free89}
	\end{figure}

We first compute the energy of three fermions ($M=3$)
for the $8\times 9$ lattice with $\PP=0$. 
For this example calculation, we have chosen 
$L_x\ne L_y$ to reduce the number of degeneracies
in the noninteracting spectrum. In 
Fig.~\ref{fig-free89}
we show the lowest five noninteracting levels and 
the corresponding states in momentum space.

\begin{figure}[ht]
	\centering
	\includegraphics[width=0.5\linewidth]{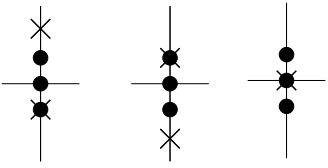}
	\caption{
Momentum space exclusions in t-matrix $M=3$ calculation
for the state (0,0)(0,1)(0,-1). The crosses
indicate exclusions when calculating pair energy for (0,0)(0,1)
(left figure), (0,0)(0,-1) (middle), and (0,1)(0,-1) (right)
respectively.
	}
	\label{fig-T89exclude}
	\end{figure}

Let us consider the lowest noninteracting state 
in the $8\times 9$, $\PP=(0,0)$, and $M=3$ system,
with three momentum vectors:
$\qq_1=(0,1)$, $\qq_2=(0,0)$, and $\qq_3=(0,-1)$
(see 
Fig.~\ref{fig-free89}).
And let us first consider the interaction of the 
pair $\qq_1$ and $\qq_2$.
The noninteracting energy
of the pair is $E^{12}_0=\E(\qq_1)+\E(\qq_2)=-7.682507$
and the total momentum is $\PP_{12}=\qq_1+\qq_2=(0,1)$.
As usual, we use $E^{12}_0$ and $\PP_{12}$ to form 
the set $\QA^{12}$ (Eq.~(\ref{eq-Q12})). Here there are no other degenerate
vectors so $\QA^{12}=\{(0,0),(0,1)\}$.
The three-particle problem
is different from the two-particle case
in the choice of $\QB^{12}$, the set of 
momentum vectors that the two particles can scatter
into. Due to the presence of the third particle and
Pauli exclusion, the two particles at $\qq_1=(0,1)$
and $\qq_2=(0,0)$ cannot be scattered into the momentum vector
$\qq_3=(0,-1)$, so we must exclude $\qq_3$
from $\QB^{12}$. Furthermore, even though
there is no particle at $\PP_{12}-\qq_3=(0,1)-(0,-1)=(0,2)$,
this momentum cannot be scattered into,
because otherwise the other particle would 
be scattered into the occupied $\qq_3$.
That is to say, the momentum vectors that can be
scattered into are 
   \begin{equation}
   \label{eq-QBthree}
   \QB^{12}=\{\qq|\qq\ne\qq_1,\qq_2,\qq_3,\PP_{12}-\qq_3\}
   \end{equation}
This exclusion is shown graphically in Fig.~\ref{fig-T89exclude}.

The t-matrix formalism can then be applied using $\QA^{12}$ and 
$\QB^{12}$ to compute energy correction $\Tt^{12}(E_{12})$
for the interaction of the $\qq_1$ and $\qq_2$ pair. 
Here $\Tt^{12}(E)$ here is the ``fermion'' function $T_{1,-1}(E)$
(Eq.~(\ref{eq-N12})),
corresponding to the antisymmetric eigenvector
of the t-matrix ${\tilde \Tm}(E)$;
the tilde denotes the modification due to exclusion
of the set $\QB^{12}$.
When the t-matrix contributes a small correction, it
is accurate to use the bare values, 
$E^0_{ij}\equiv \E(\qq_i)+\E(\qq_j)$, and this
approximation was used for all tables and figures
in this section.


The total energy within this approximation is a sum of the
t-matrix corrections for all
possible pairs in the system, which are $(\qq_2,\qq_3)$,
and $(\qq_1,\qq_3)$ in the present case:
	\begin{equation}
         E_{\rm tm}=\E(\qq_1)+\E(\qq_2)+\E(\qq_3)
        +\Tt^{12}(E_{12})+\Tt^{13}(E_{13})+\Tt^{23}(E_{23}). 
        \label{eq-Ethreefermion}
	\end{equation}
This is a special case of the effective Hamiltonian 
Eq.~(\ref{eq-multiTM}), which reduces to a $1\times 1$ matrix in the 
nondegenerate case.  (That is, whenever the set $\QA$ of
multi-fermion occupations has just one member.)
The momentum space exclusions due to the
presence of other particles are depicted in Fig.~\ref{fig-T89exclude},
and the numerical values of this calculation
are given in Table~\ref{t-T893}.

A more accurate approximation is to enforce 
a self-consistency, 
   \begin{equation}
    E_{ij} \equiv  E_{ij}^0 + \Tt^{ij}(E_{ij})
   \label{eq-selfconsistent}
   \end{equation}
where as defined above $E_{ij}^0 \equiv \E(\qq_i)+\E(\qq_j)$. 
It should be cautioned that the physical justification is
imperfect: if we visualize this approximation via a path 
integral or a Feynman diagram, the self-consistent formula 
would mean that other pairs
may be scattering simultaneously with pair $(ij)$,   yet we did not take
into account that the other pairs' fluctuations would 
modify the set of sites $\QB^{ij}$  accessible to this pair. 
In any case, analogous to the two-particle t-matrix
(Sec.~\ref{sec-tmat}), we could solve Eq.~(\ref{eq-selfconsistent}) iteratively setting
$E_{ij}^{n+1} =  E_{ij}^0 +\Tt^{ij}(E_{ij}^{n})$, 
until successive iterates agree within a tolerance
that we chose to be $10^{-15}$, which happened after some tens
iterations. 

	\begin{table}[ht]
        \centering
        \caption{
T-matrix calculation for the
$8\times 9$ lattice with $M=3$ 
noninteracting particles $\qq_1=(0,0)$, 
$\qq_2=(0,1)$, and $\qq_3=(0,-1)$.
The total noninteracting energy is 
$E_0=\E(\qq_1)+\E(\qq_2)+\E(\qq_3)$
and the total t-matrix correction is
$\Tt=\Tt^{12}+\Tt^{13}+\Tt^{23}$.
The energy calculated using the t-matrix is then
$E_{\rm tm}=E_0+\Tt$ and the exact
energy from diagonalization is $E_{\rm exact}$. 
$E^{ij}_0=\E(\qq_i)+\E(\qq_j)$, is
the noninteracting energy of the $(i,j)$ pair.
	}
\begin{ruledtabular}
        \begin{tabular}{ccrr}
$\QA^{ij}$ & $\PP_{ij}$ & $E^{ij}_0$ & $\Tt^{ij}$ 
\\\hline\
(0,0)(0,1) & (0,1)&-7.532088886&0.041949215\\
(0,0)(0,-1) & (0,-1) &-7.532088886 &0.041949215\\
(0,1)(0,-1) & (0,0) &-7.064177772 &0.118684581\\\hline
\multicolumn{2}{l}{Column sum} & 
    \multicolumn{2}{r}{$\Tt=0.202583012$} \\
\multicolumn{2}{l}{Noninteracting total} &	
    \multicolumn{2}{r}{$E_0=-11.064177772$}	 \\
\multicolumn{2}{l}{T-matrix total} &              
    \multicolumn{2}{r}{$E_{\rm tm}=-10.861594761$}\\
\multicolumn{2}{l}{Exact total} & 
    \multicolumn{2}{r}{$E_{\rm exact}=-10.871031687$} \\
        \end{tabular}
\end{ruledtabular}
        \label{t-T893}
	\end{table}

\begin{figure}[ht]
	\centering
	\includegraphics[width=\linewidth]{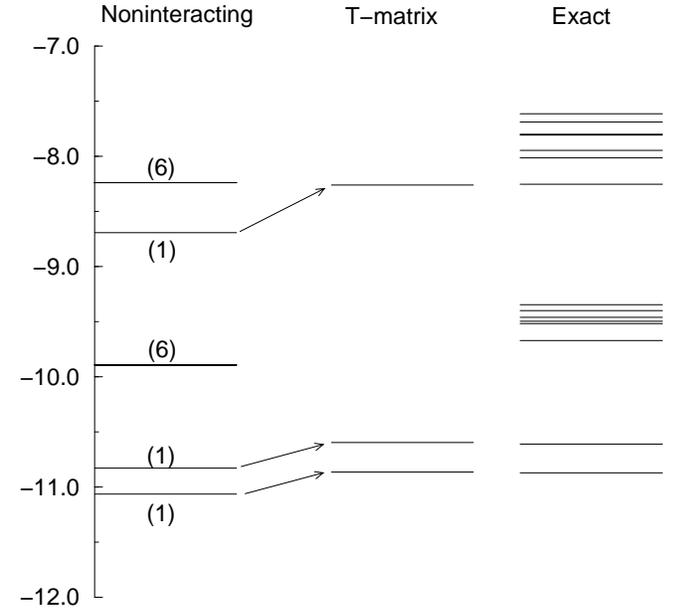}
	\caption{
Noninteracting, t-matrix, and exact energies of the three-particle
fermion system on the $8\times 9$ lattice
with $\PP=(0,0)$. The bracketed numbers
refer to the degeneracies of the level (see 
Fig.~\ref{fig-free89}).
The arrows associate the noninteracting states with the t-matrix
results. We have worked on nondegenerate noninteracting states
so far.
	}
	\label{fig-893levels}
	\end{figure}

Using the same procedure, we can also calculate the t-matrix
energies for the nondegenerate excited states
of the $M=3$ system in 
Fig.~\ref{fig-free89}: the
$(-1,0)(0,0)(1,0)$ and $(0,2)(0,0)(0,-2)$ states.
The results are shown in Table~\ref{t-893}. 
Fig.~\ref{fig-893levels} shows graphically
the noninteracting energy levels,
the t-matrix energies for the three nondegenerate states,
and the exact energies from diagonalization, 
and the arrows link the noninteracting
energies $E_0$ with the t-matrix results $E_{\rm tm}=E_0+\Tt$.
The agreement between $E_{\rm tm}$ and $E_{\rm exact}$ is good.

	\begin{table}[ht]
        \centering
        \caption{
Lowest 15 noninteracting, exact, and t-matrix
energies for $8\times 9$ lattice with $M=3$ and $\PP=(0,0)$. 
	}
\begin{ruledtabular}
        \begin{tabular}{ccc}
$E_0$ & $E_{\rm exact}$ & $E_{\rm tm}$ \\\hline
-11.064178 &	-10.871031687 &	-10.861594761 \\\hline
-10.828427 &	-10.608797838 &	-10.595561613 \\\hline
-9.892605 &	-9.672121352 &	 \\
-9.892605 &	-9.519017636 &	 \\
-9.892605 &	-9.497189108 &	 \\
-9.892605 &	-9.462304364 &	 \\
-9.892605 &	-9.398345108 &	 \\
-9.892605 &	-9.345976806 &	 \\\hline
-8.694593 &	-8.252919763 &	-8.210179503 \\\hline
-8.239901 &	-8.015024904 &	 \\
-8.239901 &	-7.946278078 &	 \\
-8.239901 &	-7.809576487 &	 \\
-8.239901 &	-7.800570818 &	 \\
-8.239901 &	-7.690625772 &	 \\
-8.239901 &	-7.615399722 &	
        \end{tabular}
\end{ruledtabular}
        \label{t-893}
	\end{table}

\subsection{A five-fermion t-matrix calculation}
\label{sec-fivefermion}

\begin{figure}[ht]
	\centering
	\includegraphics[width=0.3\linewidth]{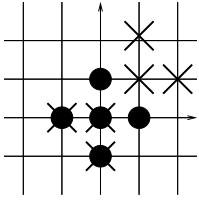}
	\caption{
Momentum space 
exclusions in t-matrix $M=5$ calculation for momentum vectors
$\qq_2=(0,1)$ and $\qq_4=(1,0)$ (dots without crosses). 
These two fermions are excluded from scattering into 
momenta from the set $\QB^{2,4}$ 
(marked by crosses). 
The ground state is shown, with occupied 
momenta (0,0), (0,1), $(0,-1)$, (1,0), and $(-1,0)$
(solid dots).
	}
	\label{fig-T895}
	\end{figure}

We now consider briefly a $M=5$ calculation, again
for the $8\times 9$ lattice. The noninteracting
ground state is unique, with momentum vectors
$\qq_1=(0,0)$, $\qq_2=(0,1)$, $\qq_3=(0,-1)$, $\qq_4=(1,0)$, 
and $\qq_5=(-1,0)$.
In Fig.~\ref{fig-T895} we show the excluded set $\QB^{2,4}$
of the t-matrix computation for the pair $(\qq_2,\qq_4)$. 
The momentum vectors $(\qq_1,\qq_3,\qq_5)$ filled with other
fermions are excluded, of course; three more wavevectors are excluded 
since the other
member of the pair would have to occupy one of $\qq_1$, $\qq_3$, 
or $\qq_5$, due to conservation of the total momentum 
$\PP = (1,1)$. 
The t-matrix results for all 10 pairs are presented
in Table~\ref{t-T895}.

One might think that the pair, $\qq_2=(0,1)$ and 
$\qq_4=(1,0)$,  
exhibits pair-exchange symmetry with (0,0)(1,1), so that 
$N_0=4$ as in Sec.~\ref{sec-N14} and Sec.~\ref{sec-4case}.   
However, since (0,0) is occupied, the (0,1)(1,0) pair 
cannot scatter into the (0,0)(1,1) pair: hence
(0,1)(1,0) is a generic pair with $N_0^{2,4}=2$. 
In general, 
if a pair is ever free to scatter into a degenerate pair state 
with a different occupation, that must be part of a many-particle
state degenerate with the original one. Thus, the complicated
t-matrix {\it pairs} with $N_0^{ij}>2$ can arise in a many-fermion
calculation only when the noninteracting {\it many-fermion} 
states are themselves degenerate.


	\begin{table}[ht]
        \centering
        \caption{
T-matrix calculation for the
$8\times 9$ lattice with five particles
(0,0), (0,1), $(0,-1)$, (1,0), and $(-1,0)$.
The exclusions in $\QB$
for the pair (0,1)(1,0) are depicted in Fig.~\ref{fig-T895}.
	}
\begin{ruledtabular}
        \begin{tabular}{lrr}
$\QA^{ij}$ & $E^{ij}_0$ & $\Tt^{ij}$ \\	\hline
$(0,0)$	$(0,1)$		&-7.532088886	&0.045184994\\
$(0,0)$	$(0,-1)$	&-7.532088886	&0.045184994\\
$(0,0)$	$(1,0)$		&-7.414213562	&0.056898969\\
$(0,0)$	$(-1,0)$	&-7.414213562 	&0.056898969\\
$(0,1)$	$(0,-1)$	&-7.064177772	&0.118684581\\
$(0,1)$	$(1,0)$		&-6.946302449	&0.081095408\\
$(0,1)$	$(-1,0)$	&-6.946302449	&0.081095408\\
$(0,-1)$ $(1,0)$	&-6.946302449	&0.081095408\\
$(0,-1)$ $(-1,0)$	&-6.946302449	&0.081095408\\
$(1,0)$	$(-1,0)$	&-6.828427125	&0.131405343\\\hline
&		$E_0=-17.892604897$	& $\Tt=0.778639481$ \\
					
		&$E_{\rm exact}=-17.145715214$&	$E_{\rm tm}=-17.113965417$\\
        \end{tabular}
\end{ruledtabular}
        \label{t-T895}
	\end{table}

\subsection{Degenerate states}
\label{sec-degstates}

In the ground state examples considered up to now
(Secs.~\ref{sec-threefermion} and \ref{sec-fivefermion}), 
the noninteracting states were all nondegenerate.
Let us now study a {\it degenerate} state in the 
third lowest level (six-fold degenerate)
of $M=3$ fermions on the $8\times 9$ lattice : 
$\qq_2=(0,1)$, $\qq_3=(1,0)$, 
$\qq_4=(-1,-1)$.
(See 
Fig.~\ref{fig-free89}, row 3.)
In this state, 
the pair $[\qq_2=(0,1),\qq_3=(1,0)]$ has the same total energy and momentum as 
the pair $[\qq_1=(0,0),\qq_5=(1,1)]$, on account of the pair 
component exchange symmetry (see Sec.~\ref{sec-setup});
consequently
$[\qq_2,\qq_3]$ {\it can} be scattered into $[\qq_1,\qq_5]$
contrary to the previous example in Sec.~\ref{sec-fivefermion}. 
Indeed, each of the six basis states in row 3 of
Fig.~\ref{fig-free89} is connected to the next one
by a two-body component exchange symmetry.

Following the two-fermion calculation with $\NA=4$
pairs (see Sec.~\ref{sec-N14}), 
the degenerate pairs $\qq_2 \qq_3$
and $[\qq_1,\qq_5]$
must be handled in the same set $\QA^{23}$. 
The results Eqs.~(\ref{eq-fstate1}), (\ref{eq-fstate2}),  
and (\ref{eq-N14f}) imply
     \begin{eqnarray}
     \T^{2,3}(E_{23})|\qq_2 \qq_3 \rangle&=&
     \frac{1}{2} (T_{1,-1,1,-1}+ T_{1,-1,-1,1}) |\qq_2 \qq_3\rangle \nonumber \\
    +\frac{1}{2}&(&T_{1,-1,1,-1}- T_{1,-1,-1,1}) |\qq_1 \qq_5\rangle .
    \label{eq-multidegen}
    \end{eqnarray}
Here 
$T_{1,-1,1,-1}$ and $T_{1,-1,-1,1}$ depend implicitly 
on $\PP=(0,0)$, on the momenta, and on 
the energy $E_{2,3}$, as well as
on $\QB^{23}$ which depends on the occupation ($\qq_4$) of the third fermion. 
In this notation, 
each $\T^{i,j}$ acting on any state produces two terms as in Eq.~(\ref{eq-multidegen}). 
The total t-matrix correction Hamiltonian is $\sum_{ij} \T^{i,j}$, summed over
all  18 possible pairs appearing in the degenerate noninteracting states. 
When we apply this to each state in the third row
of 
Fig.~\ref{fig-free89}, we finally obtain a $6\times 6$ matrix 
mixing these states. Diagonalization of this matrix would give the
correct t-matrix corrections (and eigenstates) for this
``multiplet'' of six states. 
We have not carried out such a calculation.

It is amusing to briefly consider the states in 
row 5 of 
Fig.~\ref{fig-free89}, a different sixfold degenerate set.
Unlike the row 3 case, these states separate into two
subsets of three states, of which one subset has $\{q_y\}= {-2,+1,+1}$ 
and the other subset has the opposite $q_y$ components.  
Scatterings cannot mix these subsets, so the $6\times 6$ matrix
breaks up into two identical $3\times 3$ blocks.  Hence the
t-matrix energies from row 5 consist of three twofold
degenerate levels.  By comparison, the {\it exact} interacting 
energies derived from
these noninteracting states come in three {\it nearly} degenerate pairs, 
such that the intra-pair splitting is much smaller than the 
(already small) splitting due to the t-matrix.



\subsection{Errors of the t-matrix}
\label{sec-TMerrors}

How good are the t-matrix results? From our example
calculations on the $8\times 9$ lattice,
in Tables~\ref{t-T893}, \ref{t-893}, and \ref{t-T895}, 
we see that $E_{\rm tm}$ and $E_{\rm exact}$ are close.

\begin{figure}[ht]
	\centering
	\includegraphics[width=0.70\linewidth]{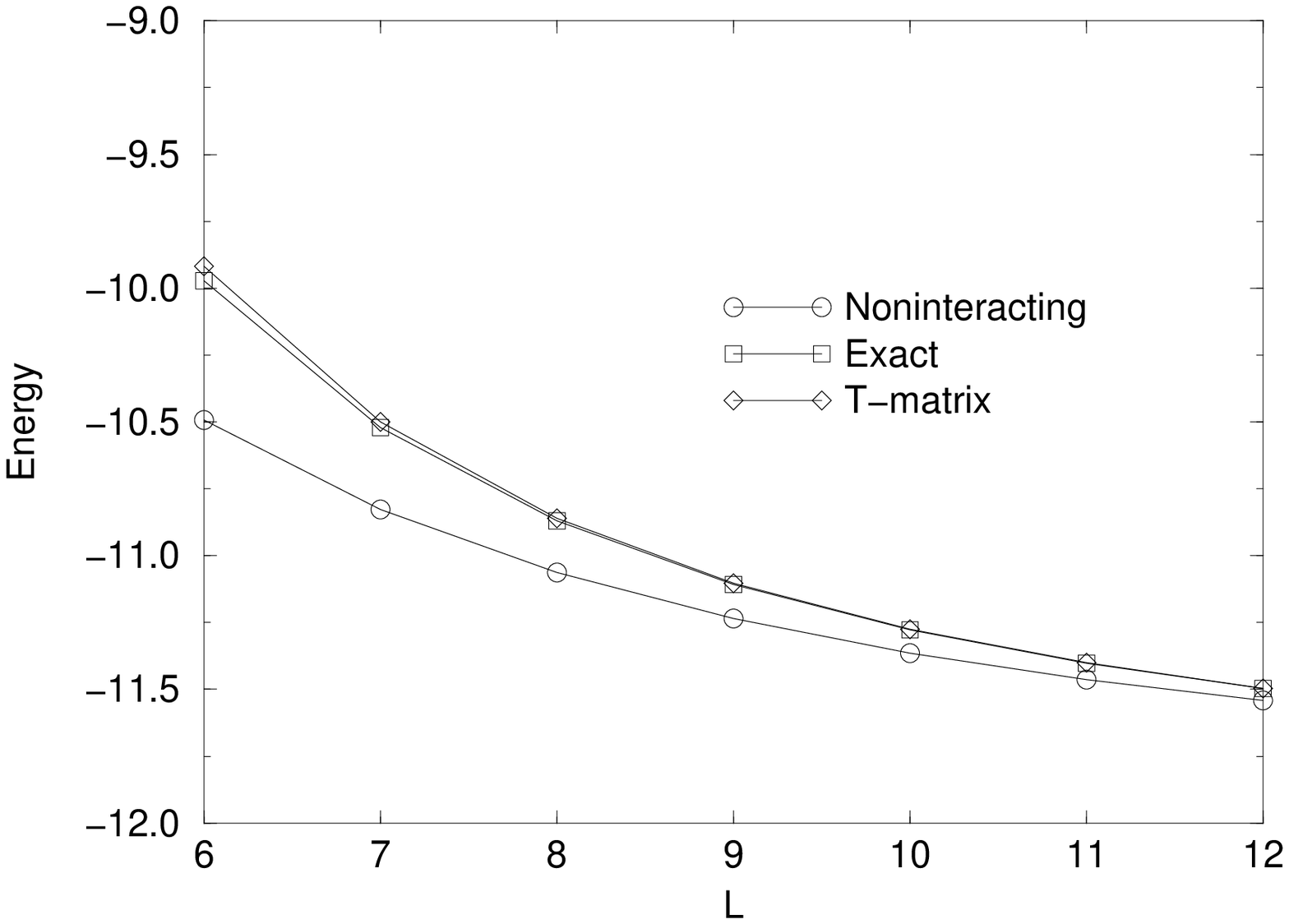}
	\includegraphics[width=0.70\linewidth]{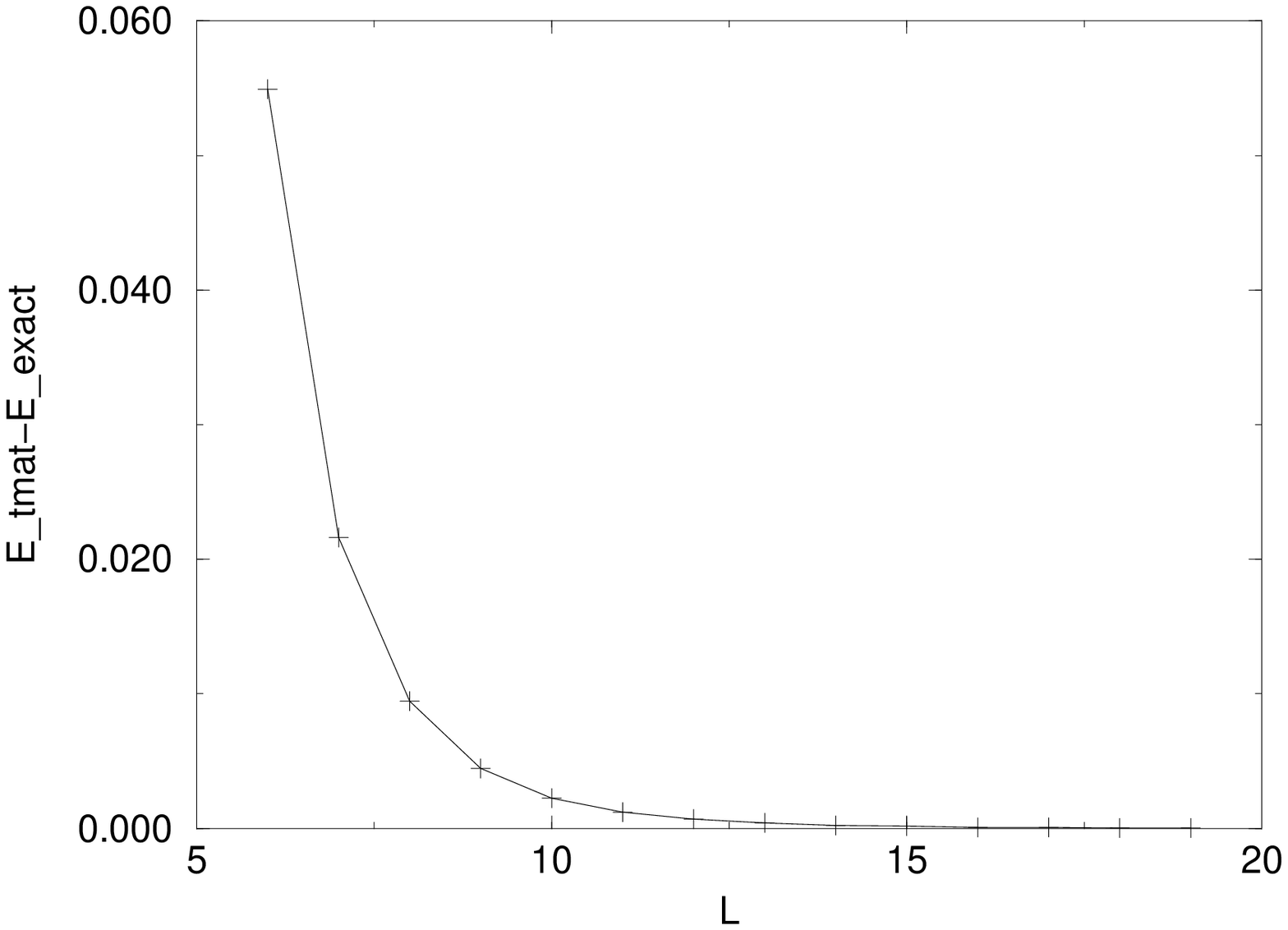}
	\caption{
Noninteracting, t-matrix, and exact energies for $M=3$,
$\PP=(0,0)$ ground state (0,0)(0,1)$(0,-1)$
on a series of $L\times (L+1)$ lattices as a function of $L$
(top graph). $E_{\rm tm}-E_{\rm exact}$ versus $L$ (bottom graph).
	}
	\label{fig-ET}
	\end{figure}

In Fig.~\ref{fig-ET} we plot the
noninteracting, t-matrix, and exact energies for
$M=3$, $\PP=(0,0)$ ground state on a series of
near square lattices $L\times (L+1)$.
The noninteracting ground state momentum vectors are
$(0,0)(0,1)(0,-1)$ for this series of lattices.
We do not plot for $L>12$, because, as can be
seen in the bottom graph, the t-matrix energy $E_{\rm tm}$ approaches
the exact energy $E_{\rm exact}$ rapidly. 
To see more clearly the error of the t-matrix result,
we plot also $E_{\rm tm}-E_{\rm exact}$, which
decays very fast as the size of the lattice increases -- 
very roughly as the $L^{-6}$ power.
Even at $L=6$, i.e. at a density $n\approx 0.05$, 
the t-matrix approximation captures 95\% of the interaction energy
$E_{\rm exact}-E_0$. 
These figures are based on using the bare energies
in $\Tt^{ij}(E_{ij})$ in Eq.~(\ref{eq-Ethreefermion}).
If we carried out the self-consistent calculation
described in Sec.~\ref{sec-threefermion}), the
error $E_{\rm tm}-E_{\rm exact}$ would be smaller by a factor of roughly 2.5. 

\section{The Dilute Limit: Energy Curves}
\label{sec-dilute}

In this section, we are interested in the
functional form of the energy as a function
of particle density for both bosons and fermions
in the dilute limit. 
In the three-dimensional case, 
the problem of dilute quantum gases with
strong, repulsive, short-range interactions
was first addressed in the language
of diagrammatic field theory by Galitskii\cite{Gal} for fermions
and Beliaev\cite{Bel} for bosons.
At that time, the ground state energy as an 
expansion in the particle density was also obtained
for hard-sphere fermion and boson gases
by Yang and collaborators~\cite{HuangYang}
using a pseudopotential method.  
The field theoretical methods were later adapted to two dimensions
in particular by Schick\cite{Schick} for 
hard-disk bosons and by Bloom\cite{Bloom} for 
hard-disk fermions.
Some other relevant analytic papers using
a t-matrix approach for the Hubbard model were 
discussed in Sec.~\ref{sec-introtmat}: 
Kanamori\cite{Kanamori} and
Mattis\cite{Mattis} in $d=3$
and Rudin and Mattis\cite{RudinMattis} 
for $d=2$. 

For both hard-disk fermions and bosons in two
dimensions, the leading-order correction to 
the noninteracting energy
is found to be in the form of $n/\ln n$, where
$n$ is particle density. Expansions with
second-order coefficients different from 
the results of Schick and Bloom were
found in Refs.~\onlinecite{BruchBoson} and \onlinecite{Hines}
for the boson case and in Refs.~\onlinecite{BruchFermion}
and \onlinecite{Engel2}
for the fermion case. There is no consensus at
this time on the correct second-order coefficient for both
the boson and fermion problems.

Recently, Ref.~\onlinecite{Lieb2D}
has proved rigorously the leading-order
expansion of the two-dimensional dilute boson
gas found by Schick.\cite{Schick}
Numerically, the dilute boson problem
on a two-dimensional lattice has been studied using quantum
Monte Carlo in Refs.~\onlinecite{Parola} and \onlinecite{Batrouni},
and they obtain good fit with Schick's result.
As we mentioned in Sec.~\ref{sec-introtmat},
more recently, because of a question regarding the validity of the
Fermi liquid theory in two dimensions 
Bloom's calculation\cite{Bloom} has received 
renewed attention,\cite{Engel2,FukuHubbard} but this result
has not been checked by numerical studies.

\subsection{Dilute bosons}
\label{sec-dilutebosons}

For two-dimensional hard disk bosons,
the energy per particle $E/M$ at the low-density limit
from diagrammatic calculations is obtained 
(in the spirit of Ref.~\onlinecite{Bel}) by Schick\cite{Schick}
	\begin{equation}
\frac{E}{M}=\frac{2\pi\hbar^2}{m}\frac{n}{|\ln(na^2)|}
\left(1+{\cal O}\left(\frac{1}{\ln(na^2)}\right)\right),
	\label{eq-Schick}
	\end{equation}
where $n=M/N$ is particle density, $m$ the mass of the boson, 
and $a$ the two-dimensional scattering length.
As mentioned above,
the coefficient of the second-order term,
has not been settled.

This hard-disk calculation was carried out
using the kinetic energy $\hbar^2 k^2/2m$. In our
lattice model, our hopping energy dispersion is (Eq.~(\ref{eq-ek})) 
	\begin{equation}
\E(\kk)=-2t(\cos(k_x)+\cos(k_y))
\approx -4t+tk^2,
	\label{eq-Etaylor}
	\end{equation}
where we have Taylor-expanded the dispersion function
near $\kk=0$ because in the dilute limit, at the ground state,
the particles occupy momentum vectors close to zero.
Therefore if we use $t=\hbar=1$ and the effective mass $m^*$
such that we have the form $\hbar^2 k^2/2m^*$, then
$m^*=1/2$ for our system. So for our model,
Schick's expansion Eq.~(\ref{eq-Schick}) should become,
	\begin{equation}
\frac{E}{M}=-4+\frac{4\pi n}{|\ln(n\ab^2)|}\left(1+
{\cal O}\left(\frac{1}{\ln(n\ab^2)}\right)\right),
	\label{eq-bosondilute}
	\end{equation}
where we have used $\ab$ to denote the 
scattering length in our lattice system.
There is no straightforward
correspondence between Schick's scattering
length $a$ in the continuum and our $\ab$ on the
lattice. With infinite nearest-neighbor repulsion,
the closest distance that our particles can come to 
is $\sqrt{2}$. We expect roughly $1 < \ab < \sqrt{2}$, 
and will determine a more precise value from curve fitting.

\begin{figure}[ht]
	\centering
	\includegraphics[width=\linewidth]{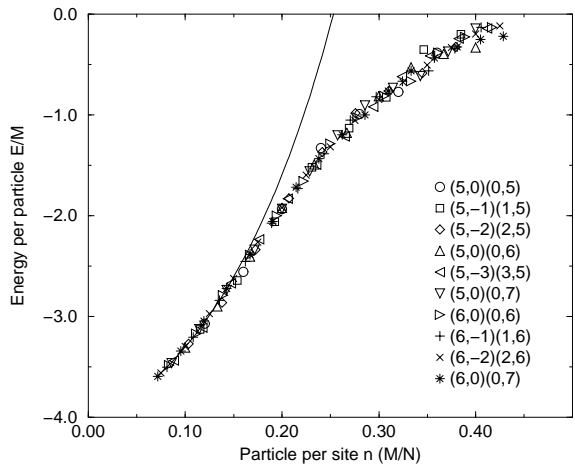}
	\caption{
Boson energy per particle $E/M$ versus particle density $M/N$ data
for ten lattices and $M\ge 3$. Data from different lattices
collapse onto one curve. The solid line corresponds to
the fitting function $-4+4\pi n(A+B/|\ln(n\ab^2)|)$ with 
$\ab=1.36$, $A=-0.016$, and $B=0.959$, which
is Eq.~(\ref{eq-bosonfit}) with parameters from
Table~\ref{t-slopes}.
	}
	\label{fig-benergycurve}
	\end{figure}

In Fig.~\ref{fig-benergycurve} we show the 
boson energy per particle ($E/M$) versus particle
per site ($M/N$) curve for ten lattices, ranging
from 25 sites to 42 sites, with three or more particles
($M\ge 3$). The data from all these lattices 
collapse onto one curve, especially
in the low-density limit.

Eq.~(\ref{eq-bosondilute}), Schick's result
applied to our model, suggests the following leading order 
fitting form for $E/M$ versus $n$ at the low-density limit,
	\begin{equation}
\frac{E/M+4}{4\pi n}=A+\frac{B}{|\ln(n\ab^2)|}.
	\label{eq-bosonfit}
	\end{equation}
That is to say, if we plot $(E/M+4)/(4\pi n)$
versus $1/|\ln(n\ab^2)|$, then, if Schick is correct,
we should get a straight line, with intercept $A=0$
and slope $B=1$, with one adjustable parameter $\ab$.

\begin{figure}[ht]
	\centering
	\includegraphics[width=\linewidth]{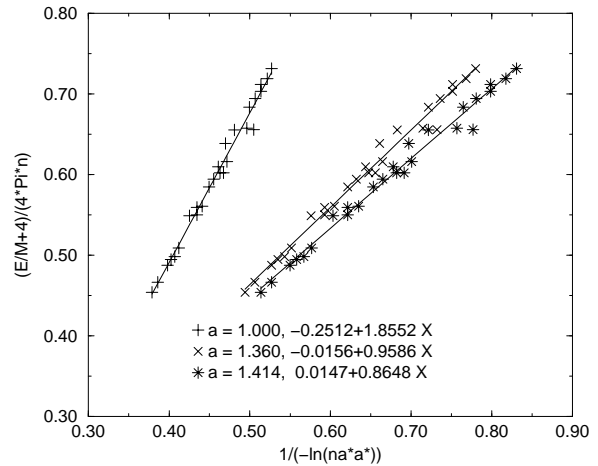}
	\caption{
$(E/M+4)/(4\pi n)$ versus $1/|\ln(n\ab^2)|$ plot to check
Schick's formula for two-dimensional dilute bosons 
(Eq.~(\ref{eq-bosondilute})). 
The data points are for $M\ge 3$ and $n\le 0.15$
from those in Fig.~\ref{fig-benergycurve}, for lattices
from $5\times 5$ to $6\times 7$. For the three $\ab$ values,
the $\ab=1.36$ choice gives 
$A=-0.016\approx 0$ and $B=0.959\approx 1$.
	}
	\label{fig-bosondilute}
	\end{figure}

In Fig.~\ref{fig-bosondilute}, we plot $(E/M+4)/(4\pi n)$
versus $1/|\ln(n\ab^2)|$ for the low-density limit
($n\le 0.15$) for three choices of $\ab=1.0, 1.36, \sqrt{2}$. 
The data points appear to lie on straight lines.
For $\ab=1.36$ the fitted intercept is $A=-0.016$
and the slope $B=0.959$. In Table~\ref{t-slopes} we
show the fitted slope and intercept for a number
of $\ab$ choices. The slope is zero close to $\ab=1.34$
and the intercept is zero close to $\ab=1.39$.
Our data thus suggest $\ab=1.36\pm 0.03$.

\begin{table}[ht]
        \centering
        \caption{
Intercept $A$ and slope $B$
in linear fitting $(E/M+4)/(4\pi n)$ versus 
$1/|\ln(n\ab)|$ for bosons, using Eq.~(\ref{eq-bosonfit}). 
The slope is one close to $\ab=1.33$
and the intercept is zero close to $\ab=1.39$.
So we get $\ab=1.36\pm 0.03$.
The fitting for three choices of $\ab$ is plotted in 
Fig.~\ref{fig-bosondilute}.
}
\begin{ruledtabular}
        \begin{tabular}{ccc|ccc|ccc} 
$\ab$ &	$B$ &	$A$& $\ab$ & $B$ & $A$ &$\ab$ & $B$ & $A$ \\\hline
1.00 &	1.855	 &-0.251 &1.32 & 1.033 & -0.039 &1.37 &	0.941 &	-0.0099\\
1.10 &	1.547 &	-0.178 &1.33 &	1.014 &	-0.033 &1.38 &	0.923 &	-0.0043\\
1.20 &	1.289 &	-0.112 &1.34	 &0.995 & -0.027&1.39 &	0.906	 &0.0013\\
1.30 &	1.072 &	-0.050 &1.35 &	0.977 &	-0.021&1.40	 &0.889 & 0.0069\\ 
1.31 &	1.053 &	-0.044 &1.36 &	0.959 &	-0.016&1.414 &	0.865 &	0.015
	\end{tabular}
\end{ruledtabular}
        \label{t-slopes}
\end{table}

In Fig.~\ref{fig-benergycurve}, the solid line
is the function $-4+4\pi n(A+B/|\ln(n\ab^2)|)$
using $\ab=1.36$, $A=-0.016$, and $B=0.959$, and
we obtain a good fit up to $n=0.15$.

For bosons, quantum Monte Carlo can be used to
obtain zero temperature energies for reasonably
large systems. For a dilute boson gas on a square lattice
with on-site hardcore but not nearest-neighbor interaction,
Ref.~\onlinecite{Parola} has fitted the 
first term of Schick's formula Eq.~(\ref{eq-Schick}),
and Ref.~\onlinecite{Batrouni} has used higher-order terms
and included the fitting of the chemical potential also.
The agreement is good in both studies.

\subsection{Dilute fermions}
\label{sec-dilutefermions}

For fermions, it customary to write the energy per particle
expansion in terms of the Fermi wavevector $k_F$.
For two-dimensional dilute hard disk fermions
with a general spin $s$, the energy per particle 
from diagrammatic calculations, is obtained by Bloom\cite{Bloom}
	\begin{equation}
\frac{E}{M}=
\frac{\hbar^2 k_F^2}{4m}\left(
1+2s\frac{1}{|\ln(k_F a)|}+{\cal O}\left(\frac{1}{\ln(k_F a)}\right)^2\right),
	\label{eq-Bloom2}
	\end{equation}
(see Ref.~\onlinecite{Bloom} for the spin-1/2 calculation
and Ref.~\onlinecite{BruchFermion} for general $s$). 

Eq.~(\ref{eq-Bloom2}) means
that for our spinless fermions ($s=0$), the leading order
correction to the noninteracting energy 
in Eq.~(\ref{eq-Bloom2}) is zero, which is due to the fact that
Eq.~(\ref{eq-Bloom2}) is derived for s-wave scattering. 
In our model, without spin,
only antisymmetric spatial wavefunctions are allowed for fermions,
and therefore the leading-order correction to the noninteracting
energy should be from p-wave scattering.
Ref.~\onlinecite{Fetter} contains a formula for
p-wave scattering in three dimensions where the
leading-order correction to $E-E_0$ is proportional
to $(k_F a)^3$ while the s-wave correction is proportional
to $k_F a$. We are not aware of a two-dimensional p-wave
calculation in the literature,\cite{FN-Bruch}
and we have not worked out this p-wave problem
in two dimensions. We expect that 
the p-wave contribution to energy should be considerably 
smaller than that from the s-wave term. 
In Sec.~\ref{sec-afewparticles}, we have considered
the case of a few fermions on a large $L\times L$
lattice, and in Fig.~\ref{fig-BFlargeL}
we have studied the interaction correction
to the noninteracting energy $\Delta E$.
It was shown there that $\Delta E$ for 
our spinless fermions is much smaller than
that for bosons.

Using $k_F^2 = 4\pi n/(2s+1)$, we can rewrite 
Eq.~(\ref{eq-Bloom2}) as
	\begin{equation}
\frac{E}{M}=
\frac{\pi \hbar^2 n}{(2s+1)m}\left(
1+4s\frac{1}{|\ln(n a^2)|}+{\cal O}\left(\frac{1}{\ln(n a^2)}\right)^2\right),
	\label{eq-Bloom2n}
	\end{equation}
In this form, it is revealed that the second term
of Eq.~(\ref{eq-Bloom2n}) is identical to the first term of the
boson expression Eq.~(\ref{eq-Schick}), apart from the replacement
$n \to 2sn/(2s+1)$. In other words, the dominant interaction
term for spinfull fermions is {\it identical} to the ferm for bosons, 
provided we replace $n$ by the density of all spin species but one, i.e.
of the spin species which can s-wave scatter off a given test
particle.

\section{Conclusion}

We have studied a two-dimensional model
of strongly-interacting fermions and bosons.
This model is the simplest model of correlated 
electrons. It is very difficult to study two-dimensional 
quantum models with short-range kinetic and
potential terms and strong interaction. There are 
very few reliable analytical methods, and many
numerical methods are not satisfactory.
With our simplified model of spinless fermions
and infinite nearest-neighbor repulsion, we can
use exact diagonalization to study systems
much larger (in lattice size) than that can
be done with the Hubbard model. One
of our goals is to publicize this model
in the strongly-correlated electron community.

In this paper, we made a systematic study of the dilute
limit of our model, using a number of analytical techniques
that so far have been scattered in the literature. 
We studied the two-particle problem
using lattice Green functions, and we demonstrated 
the use the lattice symmetry and Green function recursion
relations to simplify the complications brought
by nearest-neighbor interactions.
We derived in detail the two-particle t-matrix
for both bosons and fermions, and we showed
the difference between the boson and fermion cases
and that for fermions the first t-matrix
iteration is often a good approximation.
We applied the two-fermion t-matrix to the 
problem of a few fermions, with modifications
due to Pauli exclusion, and showed that the t-matrix
approximation is good for even small lattices.

It is somewhat puzzling that with the essential
role the t-matrix plays in almost every calculation
in the dilute limit with strong interactions,
no systematic study of the t-matrix for a lattice
model has been made, as far as we know.
We believe that our work on the two-particle t-matrix 
and the few-fermion t-matrix is first such study. 
Some approximations that
are routinely made in t-matrix calculations
are graphically presented, especially the use
of first t-matrix iteration in calculating
fermion energy. And we demonstrate
the qualitative difference between the boson 
and fermion t-matrices. We believe that this
study is a solid step in understanding dilute fermions in two dimensions,
and is of close relevance to the 2D Fermi liquid question.

The dilute boson and fermion energy per
particle curves were studied in Sec.~\ref{sec-dilute}. 
The boson curve was fitted nicely with a previous diagrammatic 
calculation, and our work on dilute bosons complements
quantum Monte Carlo results.\cite{Batrouni}
For the fermion problem in our model,
the leading order contribution to energy
is from p-wave scattering; therefore, the series
of results based on s-wave calculations
by Bloom,\cite{Bloom} Bruch,\cite{BruchFermion}
and Engelbrecht, {\it et al.}\cite{Engel2}
are not directly available. Hopefully,
the work in progress on p-wave scattering will
be completed, and our diagonalization data
can shed light to the interesting problem
of two dimensional dilute fermions.

Our model of spinless fermions and hardcore bosons with 
infinite nearest-neighbor repulsion involves a significant
reduction of the size of the Hilbert space as compared to
the Hubbard model. This enables us to obtain exact
diagonalization results for much larger lattices than
that can be done with the Hubbard model,
and this also enables us to check the various 
analytical results (Green function, t-matrix,
diagrammatics) in the dilute limit with diagonalization
for much larger systems than that has been done in previous works. 
This paper and a companion paper\cite{stripepaper}
on the dense limit are the first systematic study
of the spinless fermion model in two dimensions.
We hope that the comprehensiveness of this paper 
can not only draw more attention to this so far
basically overlooked model but also serve as 
a guide for diagonalization and analytical studies 
in the dilute limit.

\acknowledgments
This work was supported by the National Science Foundation under
grant DMR-9981744.
We thank G.~S.~Atwal for helpful discussions.

\appendix

\section{Exact Diagonalization Program}
\label{sec-diag}

This section describes briefly our exact diagonalization
program. It is indebted to Refs.~\onlinecite{Leung}
and \onlinecite{Lin}, which are guides for coding
exact diagonalization in one dimension. Here we only
describe the necessary considerations in more then
one dimension, and focuses on the use translation
symmetry to reduce the problem. 

\begin{figure}[ht]
	\centering
	\includegraphics[width=0.7\linewidth]{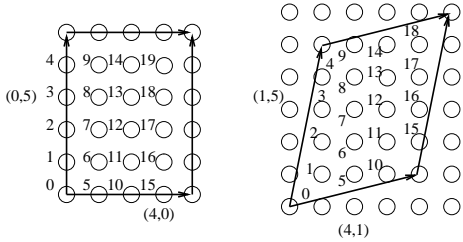}
    	\caption{Square lattices with periodic boundary conditions:
$(4,0)\times(0,5)$ on the left and $(4,1)\times(1,5)$ on the right.
Site numbers are shown, following the numbering convention,
upward and rightward.}
    	\label{fig-2lattice}
\end{figure}

Our underlying lattice is the square lattice, and we take
the lattice constant to be unity. The periodic boundary
conditions are specified by two lattice vectors
${\bf R}_1$ and  ${\bf R}_2$, such that
for any lattice vector ${\bf r}$ we have
${\bf r}+n_1{\bf R}_1+n_2{\bf R}_2\equiv{\bf r}$,
where $n_1$ and $n_2$ are two integers.
In Fig.~\ref{fig-2lattice}, we show two systems. The first
one has ${\bf R}_1=(4,0)$ and ${\bf R}_2=(0,5)$
so the number of lattice sites is $N=20$. The
second one has ${\bf R}_1=(4,1)$ and ${\bf R}_2=(1,5)$
so $N=|{\bf R}_1\times{\bf R}_2|=19$.
 From this example we see immediately the advantage of 
having skewed boundary conditions: we can have reasonably
shaped systems with number of sites (here 19) not possible for
an usual rectangular system.

A site order is needed to keep track of the order
of the fermion sign. The convention 
that we use is starting the zeroth site from the
lower left corner and move progressive upward and rightward
following the square lattice structure until we encounter
boundaries of the lattice defined by 
the periodic boundary condition vectors 
${\bf R}_1$ and ${\bf R}_2$ (see Fig.~\ref{fig-2lattice}).
A basis state with $M$ particles is then represented
by an array of the $M$ occupied site numbers,
with nearest neighbors excluded (because $V=+\infty$
in our Hamiltonian Eq.~(\ref{eq-Ham})).
Denote such a basis state $|n\rangle$
and we have 
	\begin{equation}
	H|n\rangle=-t\sum_{m\in {\cal M}} s_m |m\rangle,
	\label{eq-Hn}
	\end{equation}
where ${\cal M}$ denotes the set of states 
created by hopping one particle in $|n\rangle$ 
to an allowed nearest-neighbor site
and for bosons $s_m=1$ always and for fermions
$s_m=\pm 1$.\cite{FN-sm} And the matrix
element is $\langle m | H|n\rangle = -s_m t$ 
if $m\in {\cal M}$ and 0, otherwise.

In order to calculate for large systems,
it is necessary to use symmetry to block diagonalize
the Hamiltonian matrix. In our code, we use lattice
translation symmetry because it works
for arbitrary periodic boundaries. The eigenstate
that we use is the Bloch state\cite{Leung}
	\begin{equation}
	|n{\bf k}\rangle=\frac{1}{N_{n{\bf k}}}\sum_{l=0}^{N-1} 
	e^{-i{\bf k}\cdot{\bf R}_l} T_l |n\rangle.
	\label{eq-Bloch}
	\end{equation}
 In this expression ${\bf k}$ is a wavevector 
(one of $N$, where $N$ is the number of sites),
$\RR_l$ is a lattice vector
$T_l$ is a short hand notation for translation by ${\bf R}_l$,
and $N_{n{\bf k}}$ is a normalization factor. 
The original basis states are divided by translation
into classes and any two states in the same class
give the same Bloch state with an overall phase factor. What we 
need to do is to choose a representative from each class,
and use this state consistently to build Bloch states.
For a state $|n\rangle$ we denote its representative 
$|\bar n\rangle$. 

To compute the Hamiltonian matrix elements using
the Bloch states Eq.~(\ref{eq-Bloch}),
let us start from a representative state $|\bar n\rangle$.
We have, as in Eq.~(\ref{eq-Hn}),
$H|\bar n\rangle=-t\sum_{m\in{\cal M}} s_m |m\rangle$, 
then $H|T_l \bar n\rangle=T_l H|\bar n\rangle
=-\sum_m s_m T_l |m\rangle$, where we have used
the fact that $T_l$ commutes with the Hamiltonian.
We have,
	\begin{eqnarray}
	\nonumber
&&H|\bar n{\bf k}\rangle=
\frac{1}{N_{\bar n{\bf k}}}\sum_{l=0}^{N-1} 
e^{-i{\bf k}\cdot{\bf R}_l} H T_l |\bar n\rangle\\
	\nonumber
&&=-\frac{1}{N_{\bar n{\bf k}}}\sum_{m\in{\cal M}} s_m
\sum_{l=0}^{N-1} 
e^{-i{\bf k}\cdot{\bf R}_l} T_l |m\rangle\\
&&=-\frac{1}{N_{\bar n{\bf k}}}\sum_{m\in{\cal M}}N_{m{\bf k}} s_m 
|m{\bf k}\rangle.
	\end{eqnarray}
Next because we are interested in matrix
elements between representative states, we want to
connect $|m{\bf k}\rangle$ in the preceding 
equation to $|\bar m{\bf k}\rangle$.
If $T_{j(m)}|\bar m\rangle=\sigma_{j(m)} |m\rangle$,
then $|m{\bf k}\rangle
=\sigma_{j(m)} e^{i{\bf k}\cdot{\bf R}_{j(m)}} |\bar m{\bf k}\rangle$.
So we have 
	\begin{equation}
H|\bar n{\bf k}\rangle=
-\frac{1}{N_{\bar n{\bf k}}}\sum_{m\in{\cal M}}N_{\bar m{\bf k}} s_m
\sigma_{j(m)} e^{i{\bf k}\cdot{\bf R}_{j(m)}} |\bar m{\bf k}\rangle.
	\label{eq-Hnk}
	\end{equation}
We should note that for all $m\in{\cal M}$ there can be
more than one element having the same representative $|\bar m\rangle$.
That is to say in the sum in Eq.~(\ref{eq-Hnk}),
there can be more than one term with $|\bar m{\bf k}\rangle$.
We write a new set ${\cal M}'=\{m|m\in{\cal M} \mbox{ and $m$
has rep }\bar m\}$. Then we can write our matrix
element equation as follows,
	\begin{equation}
\langle \bar m{\bf k}|H|\bar n{\bf k}\rangle
=-\frac{N_{\bar m{\bf k}}}{N_{\bar n{\bf k}}}
\sum_{m\in{\cal M}'}\sigma_{j(m)} e^{i{\bf k}\cdot{\bf R}_{j(m)}}s_m.
	\label{eq-Hmn}
	\end{equation}
Eq.~(\ref{eq-Hmn}) is the centerpiece of the Bloch state calculation.
It includes many of the complications that come with the Bloch basis set.
(See Ref.~\onlinecite{Leung} for the corresponding equation
in one dimension.)

Let us use $\N$ to denote the number of Bloch basis states
for one $\kk$. $\N$ is the dimensionality of the matrix that
we need to diagonalize. For $\N$ in the order of thousands,
full diagonalization (with storage of the matrix)
is done using LAPACK,\cite{LAPACK} and a $3156\times 3156$ matrix 
($7\times 7$ with $M=18$) takes about 27 minutes.\cite{FN-machine}
For larger $\N$, the Lanczos method is implemented following
the instructions in Ref.~\onlinecite{Leung},
We have two options. First, we store information 
about the matrix (i.e., for each
column, a set of ($p$, $j(m)$, $\sigma_{j(m)}$)
described above that
contains information about the nonzero entries
of the Hamiltonian matrix in this column).
The $M=9$ case on $7\times 7$ lattice with $\N=1,120,744$
and tolerance $10^{-15}$ takes about 45 minutes (32
Lanczos iterations) and uses about 1.5 GB of memory. This 
basically reaches our memory limit. 

On the other hand, we can also do 
Lanczos without storing matrix information. 
The same $M=9$ case on $7\times 7$ uses
only 200 MB of memory but takes
more than four hours (263 minutes),
for a larger tolerance $10^{-7}$ (therefore fewer
Lanczos iterations, 14). 
Without storing matrix information, we can calculate
for larger matrices: the $M=11$ case on $7\times 7$, 
with $\N=1,906,532$ (the largest for the 
$7\times 7$ system) and tolerance $10^{-7}$, 
is done in 10 hours, using less than
400 MB of memory. The largest matrix we computed
for this work is $\N=2,472,147$,
i.e., about 2.5 million Bloch states, for
$M=4$ on $20\times 20$. This takes 10 hours
and uses about 550 MB of memory, for
a tolerance of $10^{-7}$.

In addition, we have also installed ARPACK\cite{ARPACK}
that uses the closely related so-called Arnoldi methods 
and can obtain excited state eigenvalues and eigenvectors as well.
If we only need information about the ground state,
our Lanczos program is considerably faster than ARPACK.

The $7\times 7$ lattice, with maximum $\N$ around 2 million,
is basically the largest lattice for which we can calculate
eigenenergies at all fillings. The exponential growth 
is very rapid after this.
The $8\times 8$ lattice with 8 particles has $9,151,226$ 
Bloch states, and with one more particle, 
$M=9$, there are $30,658,325$, 
i.e., more than 30 million states.

\section{Physical meaning of $\Tm(E)$}
\label{sec-physical}

In this section we give yet another derivation
of the t-matrix which makes more explicit
the physical meaning of $T(E,\PP;\qq,\qq')$ 
Eq.~(\ref{eq-Gq}).

\begin{figure}[ht]
	\centering
	\includegraphics[width=0.8\linewidth]{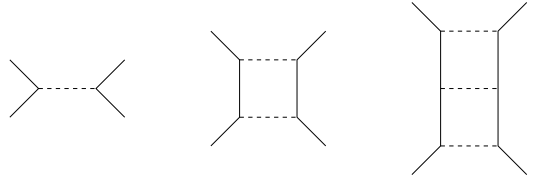}
\caption{
The three figures represent perturbative terms
involving $V(\qq-\qq')$,
$T_2(E,\PP;\qq,\qq')$ and $T_3(E,\PP;\qq,\qq')$.
The t-matrix, $T(E,\PP;\qq,\qq')$, is
the sum of all these terms, i.e., it is the sum
of the ladder diagrams to infinite order.
}
	\label{fig-graph}
	\end{figure}

Before we get into a lot of algebra, let us
describe the physical idea. In scattering theory
we know that the Born series is a perturbation
series of the scattering amplitude in terms
of the potential. In Fig.~\ref{fig-graph} we show
the first three terms graphically,
where the first term, the first
Born approximation, is particularly simple--it is
the Fourier transform of the potential.
We also know that when the potential is weak
the first few terms are an good approximation
to the scattering amplitude, but when the
potential is strong, we need all terms.
In this section, we will show that our t-matrix
$T(E,\PP,\qq,\qq')$ is the sum of all such
two-body scattering terms.

We start with Eq.~(\ref{eq-sch}) which we copy here for convenience,
	\begin{equation}
	\label{eq-1}
(E-\E(\qq)-\E(\PP-\qq))g(\qq)=\frac{1}{N}\sum_{\qq'} V(\qq-\qq')g(\qq').
	\end{equation}
For $\qq\in \QA$ we break up the sum over $\qq'$ into two terms
and get,
	\begin{eqnarray}
\nonumber
(E-E_0)g(\qq)&=&\frac{1}{N}\sum_{\qq'\in \QA} V(\qq-\qq')g(\qq')\\
&+&\frac{1}{N}\sum_{\qq'\in \QB} V(\qq-\qq')g(\qq').
	\label{eq-2}
	\end{eqnarray}
For $\qq\in \QB$ we can rewrite Eq.~(\ref{eq-1}) to get,
	\begin{equation}
	\label{eq-3}
g(\qq')=\frac{1}{N}\sum_{\qq''}\frac{V(\qq'-\qq'')}
{E-\E(\qq')-\E(\PP-\qq')}g(\qq'').
	\end{equation}
Plug Eq.~(\ref{eq-3}) into the second sum in Eq.~(\ref{eq-2}) and rearrange
terms, we get,
	\begin{eqnarray}
\nonumber
(E-E_0)g(\qq)&=&\frac{1}{N}\sum_{\qq'\in \QA} V(\qq-\qq')g(\qq')\\
&+&\frac{1}{N}\sum_{\qq''} T_2(E,\PP;\qq,\qq'')g(\qq''),
	\label{eq-4}
	\end{eqnarray}
where we have defined,
	\begin{equation}
	\label{eq-5}
T_2(E,\PP;\qq,\qq'')=\frac{1}{N}\sum_{\qq'\in \QB}
\frac{V(\qq-\qq')V(\qq'-\qq'')}{E-\E(\qq')-\E(\PP-\qq')}
	\end{equation}
Now break the sum over $\qq''$ in Eq.~(\ref{eq-4}) into two
parts, and we get
	\begin{eqnarray}
\nonumber
(E-E_0)g(\qq)&=&\frac{1}{N}\sum_{\qq'\in \QA} V(\qq-\qq')g(\qq')\\
\nonumber
&+&\frac{1}{N}\sum_{\qq'\in \QA} T_2(E,\PP;\qq,\qq')g(\qq')\\
&+&\frac{1}{N}\sum_{\qq'\in \QB} T_2(E,\PP;\qq,\qq')g(\qq').
	\label{eq-6}
	\end{eqnarray}
Plug Eq.~(\ref{eq-3}) into the last term of Eq.~(\ref{eq-6}) and we get
	\begin{eqnarray}
\nonumber
(E-E_0)g(\qq)&=&\frac{1}{N}\sum_{\qq'\in \QA} V(\qq-\qq')g(\qq')\\
\nonumber
&+&\frac{1}{N}\sum_{\qq'\in \QA} T_2(E,\PP;\qq,\qq')g(\qq')\\
&+&\frac{1}{N}\sum_{\qq'} T_3(E,\PP;\qq,\qq')g(\qq'),
	\label{eq-7}
	\end{eqnarray}
where we have defined
\begin{widetext}
	\begin{equation}
T_3(E,\PP;\qq,\qq')
=\frac{1}{N^2}\sum_{\qq'',\qq'''\in \QB}
\frac{V(\qq-\qq'')V(\qq''-\qq''')V(\qq'''-\qq')}
{(E-\E(\qq'')-\E(\PP-\qq''))(E-\E(\qq''')-\E(\PP-\qq'''))}
	\label{eq-8}
	\end{equation}
\end{widetext}
Continue this process, we obtain
	\begin{eqnarray}
\nonumber
&&(E-E_0)g(\qq)=
\frac{1}{N}\sum_{\qq'\in \QA} (
V(\qq-\qq')\\&&+T_2(E,\PP;\qq,\qq')
+T_3(E,\PP;\qq,\qq')+...
)g(\qq').
	\label{eq-9}
	\end{eqnarray}

What we have done here is the traditional
perturbation theory using iteration.
Eq.~(\ref{eq-9}) is the Born series for scattering amplitude.
The first term $V(\qq-\qq')$,
the Fourier transform of the potential $V(\rr)$,
is the first Born approximation.
The content of higher order terms $T_2$, $T_3$, ...
can be obtained from their definition.
Eq.~(\ref{eq-5}) says that $T_2$ involves two scatterings
under $V$, and Eq.~(\ref{eq-8}) says that $T_3$ involves three
scatterings. Thus the Born series Eq.~(\ref{eq-9}) can be graphically
depicted at the ladders in Fig.~\ref{fig-graph},\cite{Holstein}
and it involves multiple scatterings to all orders.
Note that each term in the Born series is infinite
for infinite potential $V$.
Next we will show that summing all the terms in the series
gives the t-matrix and the potential $V$ cancels
out, giving a finite value.

It is easy to check that
\begin{equation}
T_2(E,\PP;\qq,\qq')=V^2\sum_{ij}
e^{i\qq\cdot\RR_i}e^{-i\qq'\cdot\RR_j} \GRB_{ij}(E,\PP),
\end{equation}
where $\GRB_{ij}(E,\PP)$ is our good old Green
function Eq.~(\ref{eq-green2}),
\begin{equation}
T_3(E,\PP;\qq,\qq')=V^3\sum_{ij}
e^{i\qq\cdot\RR_i}e^{-i\qq'\cdot\RR_j} (\GRB(E,\PP))^2_{ij},
\end{equation}
and 
\begin{equation}
T_n(E,\PP;\qq,\qq')=V^n\sum_{ij}
e^{i\qq\cdot\RR_i}e^{-i\qq'\cdot\RR_j} (\GRB(E,\PP))^{n-1}_{ij}.
\end{equation}
Now plug these results into Eq.~(\ref{eq-9}),
we get
\begin{widetext}
	\begin{equation}
(E-E_0)g(\qq)=\sum_{\qq'\in \QA}
\left[\frac{1}{N}\sum_{ij}
e^{i\qq\cdot\RR_i}e^{-i\qq'\cdot\RR_j}
V\left(\delta_{ij}+V\GRB_{ij}+V^2(\GRB_{ij})^2+...\right)\right].
\label{eq-10}
\end{equation}
\end{widetext}
Now we come to a formal step,
\begin{equation}
V\left(I+V\GRB+V^2(\GRB)^2+...\right)
=V(I-V\GRB(E))^{-1},
\label{eq-geometric}
\end{equation}
and the interesting result
is that the infinite potential $V$ cancels out,
giving a finite value $-\GRB(E)^{-1}$.

If we can do this formal sum, then we get
from Eq~(\ref{eq-10}),
\begin{eqnarray}
\nonumber
(E-E_0)g(\qq)
=\sum_{\qq'\in \QA}T(E,\PP;\qq,\qq')g(\qq'),
\end{eqnarray}
which is exactly our momentum space T-matrix
equation Eq.~(\ref{eq-Gqeq}) and $T(E,\PP;\qq,\qq')$
is exactly our t-matrix Eq.~(\ref{eq-Gq}).


\begin{thebibliography}{99}
\bibitem{FN-halffilled}
Note that with spinless fermions and hardcore bosons
and infinite nearest-neighbor repulsion, the filling (particle per 
lattice site) in our model goes from 0 to 1/2 only.
\bibitem{FN-notation}
We are slightly abusing the notation $V$. $V(\rr)$ is the potential
function for all space. $V$ is the value of the potential
at four points ($V=V(\rr=(1,0))$ for example). And $V(\kk)$ the
Fourier transform of $V(\rr)$. We will differentiate 
these meanings by including explicitly
the argument, $\rr$ or $\kk$. Without argument, we then
mean $V=V(\rr=(1,0))$.
\bibitem{FN-Blaer}
The following derivation is indebted to Ref.~\onlinecite{Blaer} which 
solved a similar problem.
\bibitem{FN-component}
Note that, anisotropic hopping dispersion
$\E(\qq)=-2t_x\cos q_x-2t_y\cos q_y$,
where $\E_x(q)=-2t_x\cos q$ 
and $\E_y(q)=-2t_y\cos q$, also
has this pair component exchange symmetry.
\bibitem{FN-Bruch}
Ref.~\onlinecite{BruchFermion} discusses spin degeneracy,
but does not contain a formula for p-wave terms.
\bibitem{FN-sm}
$s_m$ is in fact a function of $m$ and $n$, but in our discussion
in this paper, we always consider a fixed column $n$,
so we omitted the $n$ dependence in the notation $s_m$.
\bibitem{FN-machine}
The computer running times in this paper are obtained using an
Intel Pentium III 700 processor.
\end{thebibliography}
\end{document}